\documentclass[longauth]{aa}

\usepackage{txfonts}
\usepackage{graphicx}
\usepackage{natbib}
\usepackage{color}
\usepackage[breaklinks=true]{hyperref}

\usepackage{natbib,twoopt}
\bibpunct{(}{)}{;}{a}{}{,}            
\makeatletter
  \newcommandtwoopt{\citeads}[3][][]{\href{http://adsabs.harvard.edu/abs/#3}%
    {\def\hyper@linkstart##1##2{}%
     \let\hyper@linkend\@empty\citealp[#1][#2]{#3}}}
  \newcommandtwoopt{\citepads}[3][][]{\href{http://adsabs.harvard.edu/abs/#3}%
    {\def\hyper@linkstart##1##2{}%
     \let\hyper@linkend\@empty\citep[#1][#2]{#3}}}
  \newcommandtwoopt{\citetads}[3][][]{\href{http://adsabs.harvard.edu/abs/#3}%
    {\def\hyper@linkstart##1##2{}%
     \let\hyper@linkend\@empty\citet[#1][#2]{#3}}}
  \newcommandtwoopt{\citeyearads}[3][][]%
    {\href{http://adsabs.harvard.edu/abs/#3}
    {\def\hyper@linkstart##1##2{}%
     \let\hyper@linkend\@empty\citeyear[#1][#2]{#3}}}
\makeatother

\newcommand{\nwd}{14\,844}
\newcommand{\tout}{T_{\rm out}}
\newcommand{\pca}{p_{\rm Ca}}
\newcommand{\teff}{T_{\rm eff}}
\newcommand{\logg}{\log {\rm g}}

\begin{document}

\title{J-PLUS: Understanding outlier white dwarfs in the third data release via dimensionality reduction}

\author{C.~L\'opez-Sanjuan\inst{\ref{CEFCA},\ref{UA}}
\and P.-E.~Tremblay\inst{\ref{warwick}}
\and A.~del~Pino\inst{\ref{IAA}}
\and H.~Dom\'{\i}nguez S\'anchez\inst{\ref{IFCA}}
\and H.~V\'azquez Rami\'o\inst{\ref{CEFCA},\ref{UA}}
\and A.~Ederoclite\inst{\ref{CEFCA},\ref{UA}}
\and A.~J.~Cenarro\inst{\ref{CEFCA},\ref{UA}}
\and A.~Mar\'{\i}n-Franch\inst{\ref{CEFCA},\ref{UA}}
\and B.~Anguiano\inst{\ref{CEFCA}}
\and T.~Civera\inst{\ref{CEFCA}}
\and P.~Cruz\inst{\ref{CAB}}
\and J.~A.~Fern\'andez-Ontiveros\inst{\ref{CEFCA},\ref{UA}}
\and F.~M.~Jim\'enez-Esteban\inst{\ref{CAB}}
\and A.~Rebassa-Mansergas\inst{\ref{UPC},\ref{IEEC}}
\and J.~Vega-Ferrero\inst{\ref{CEFCA},\ref{UA}}
\and J.~Alcaniz\inst{\ref{ON}}
\and R.~E.~Angulo\inst{\ref{DIPC},\ref{ikerbasque}}
\and D.~Crist\'obal-Hornillos\inst{\ref{CEFCA}}
\and R.~A.~Dupke\inst{\ref{ON},\ref{MU}}
\and C.~Hern\'andez-Monteagudo\inst{\ref{IAC},\ref{ULL}}
\and M.~Moles\inst{\ref{CEFCA}}
\and L.~Sodr\'e Jr.\inst{\ref{USP}}
\and J.~Varela\inst{\ref{CEFCA}}
}

\institute{
Centro de Estudios de F\'{\i}sica del Cosmos de Arag\'on (CEFCA), Plaza San Juan 1, 44001 Teruel, Spain\\\email{clsj@cefca.es}\label{CEFCA}
\and Unidad Asociada CEFCA-IAA, CEFCA, Unidad Asociada al CSIC por el IAA y el IFCA, Plaza San Juan 1, 44001 Teruel, Spain.\label{UA}
\and Department of Physics, University of Warwick, Coventry, CV4 7AL, UK\label{warwick}
\and Instituto de Astrof\'{\i}sica de Andaluc\'{\i}a, IAA-CSIC, Glorieta de la Astronom\'{\i}a S/N, 18008 Granada, Spain\label{IAA}
\and Instituto de F\'{\i}sica de Cantabria (IFCA, CSIC), Avda. de los Castros s/n, 39005 Santander, Spain\label{IFCA}
\and Centro de Astrobiolog\'{\i}a (CAB), CSIC-INTA, Camino Bajo del Castillo s/n, 28692 Villanueva de la Ca\~nada, Madrid, Spain\label{CAB}
\and Departament de F\'isica, Universitat Polit\`ecnica de Catalunya, c/ Esteve Terrades 5, 08860 Castelldefels, Barcelona, Spain\label{UPC}
\and Institut d’Estudis Espacials de Catalunya, Esteve Terradas, 1, Edifici RDIT, Campus PMT-UPC, 08860 Castelldefels, Barcelona, Spain\label{IEEC}
\and Observat\'orio Nacional - MCTI (ON), Rua Gal. Jos\'e Cristino 77, S\~ao Crist\'ov\~ao, 20921-400 Rio de Janeiro, Brazil\label{ON}
\and Donostia International Physics Centre (DIPC), Paseo Manuel de Lardizabal 4, 20018 Donostia-San Sebastián, Spain\label{DIPC}
\and IKERBASQUE, Basque Foundation for Science, 48013, Bilbao, Spain\label{ikerbasque}
\and University of Michigan, Department of Astronomy, 1085 South University Ave., Ann Arbor, MI 48109, USA\label{MU}
\and Instituto de Astrof\'{\i}sica de Canarias, La Laguna, 38205, Tenerife, Spain\label{IAC}
\and Departamento de Astrof\'{\i}sica, Universidad de La Laguna, 38206, Tenerife, Spain\label{ULL}
\and Instituto de Astronomia, Geof\'{\i}sica e Ci\^encias Atmosf\'ericas, Universidade de S\~ao Paulo, 05508-090 S\~ao Paulo, Brazil\label{USP}
}

\date{Received 24 April 2025 / Accepted 26 July 2025}

\abstract
{}
{We present the white dwarf catalog derived from the third data release of the Javalambre Photometric Local Universe Survey (J-PLUS DR3), which covers $3\,284$ deg$^2$ using $12$ optical filters. A particular focus is given to the classification of outlier sources, those largely incompatible with the theoretical models used in the analysis, through dimensionality reduction techniques.}
{We applied a Bayesian fitting process to the $12$-band J-PLUS photometry of white dwarf candidates from {\it Gaia} EDR3. The derived parameters were effective temperature, surface gravity, and parallax. We used theoretical models from H- and He-dominated atmospheres, with priors applied to parallax and spectral type. From the posteriors, we derived the probability of an H-dominated atmosphere and of calcium absorption for each source. Outliers were identified as sources with $\chi^2 \geq 23.2$, indicating significant deviations from the best-fitting model. We analyzed the residuals from the fits using the uniform manifold approximation and projection (\texttt{UMAP}) technique, which enables the classification of outliers into distinct categories.}
{The catalog includes $\nwd$ white dwarfs with $r \leq 20$ mag and $1 \leq \varpi < 100$ mas, with $72$\% of the sources lacking spectroscopic ($R \gtrsim 500$) classification. The application of \texttt{UMAP} to the residuals identified three main types of outliers: random measurement fluctuations (391 sources), metal-polluted white dwarfs (98 sources), and two-component systems (282 sources). The last category also includes white dwarfs with strong carbon absorption lines. We validated the reliability of J-PLUS classifications by comparison with spectroscopy from the Sloan Digital Sky Survey and the Dark Energy Spectroscopic Instrument, and with {\it Gaia} BP/RP low-resolution spectra, confirming a one-to-one correspondence between J-PLUS photometric and spectroscopic classifications.}
{The J-PLUS DR3 white dwarf catalog provides a robust dataset for statistical studies. The use of dimensionality reduction techniques enhances the identification of peculiar objects, making this catalog a valuable resource for the selection of interesting targets such as metal-polluted white dwarfs or binary systems.}


\titlerunning{J-PLUS. Understanding outlier white dwarfs via dimensionality reduction}

\authorrunning{L\'opez-Sanjuan et al.}

\maketitle

\section{Introduction}\label{sec:intro}
White dwarfs represent the final evolutionary stage of low- and intermediate-mass stars ($M < 8-10~M_{\odot}$), making them key objects for understanding stellar evolution, the star formation history of the Milky Way, and advances in the physics of dense matter \citep[e.g.][]{isern22}. The {\it Gaia} mission \citep{gaia} has revolutionized the study of white dwarfs \citep{tremblay24}. The {\it Gaia} second data release (DR2; \citealt{gaiadr2}) marked a major breakthrough by identifying above $200\,000$ white dwarf candidates \citep{jimenezesteban18,gentilefusillo19}, while {\it Gaia} early data release three (EDR3; \citealt{gaiaedr3}) further refined these selections and expanded the dataset, which now contains approximately $360\,000$ candidates with precise astrometry and photometry \citep[][hereafter GF21]{GF21}. The {\it Gaia} data enable the construction of volume-limited samples \citep[e.g.,][]{hollands18,gaia_edr3_nearby,jimenezesteban23,40pciv,kilic25}, minimizing selection biases and allowing robust statistical analyses of the white dwarf population.

Although photometric and astrometric data from {\it Gaia} provide valuable information, spectroscopic follow-up remains essential to further characterize their atmospheric properties. Before {\it Gaia}, the Sloan Digital Sky Survey (SDSS; \citealt{sdss_dr17}) played a crucial role in building spectroscopic catalogs of white dwarfs, providing thousands of high-quality spectra (e.g., \citealt{eisenstein06,kepler19}; GF21). More recently, dedicated spectroscopic efforts have resulted in highly complete catalogs, such as the $40$ pc sample \citep{limoges15,40pci,40pcii,40pciii,40pciv} and the $100$ pc sample \citep{kilic20,kilic25}. Currently, massive spectroscopic follow-ups of the {\it Gaia}-based samples are ongoing or planned with the SDSS-V Milky Way Mapper \citep{sdssv}, the William Herschel Telescope Enhanced Area Velocity Explorer (WEAVE, \citealt{weave_new}), the Dark Energy Spectroscopic Instrument (DESI, \citealt{desi_mws,desi_edr_wd}), and the 4-meter Multi-Object Spectrograph Telescope (4MOST, \citealt{4most_wd}). These surveys collectively aim to observe over $250\,000$ white dwarfs spectroscopically, providing precise atmospheric parameters and composition. Additionally, the low-resolution blue photometer and red photometer (BP/RP) spectra from {\it Gaia} DR3 \citep{gaiadr3_bprp, gaiadr3_calib} also permit white dwarf classification and atmospheric characterization, further expanding the potential for statistical studies of the white dwarf population \citep[e.g.,][]{jimenezesteban23,torres23,garciazamora23,vincent24,perezcouto24}.

Another valuable complement to {\it Gaia}’s catalogs of white dwarfs comes from large-area multi-filter photometric surveys, which provide deeper observations over thousands of square degrees and enable the construction of spectral energy distributions (SEDs) without requiring any additional spectroscopic preselection. The most relevant projects in this context are the Javalambre Photometric Local Universe Survey\footnote{\url{https://www.j-plus.es}} (J-PLUS, with $12$ optical filters; \citealt{cenarro19}, Table~\ref{tab:filters}), the Southern Photometric Local Universe Survey\footnote{\url{https://datalab.noirlab.edu/splus}} (S-PLUS, with the same filter system as J-PLUS; \citealt{splus}), and the Javalambre Physics of the Accelerating Universe Astrophysical Survey\footnote{\url{https://www.j-pas.org}} (J-PAS, with 56 narrow-band filters of 14.5 nm width and continuous coverage from $350$ to $930$ nm; \citealt{jpas,minijpas}). Using its multi-filter photometric coverage, J-PLUS has already contributed significant results, such as constraining the spectral evolution of white dwarfs \citep{clsj22pda} and determining the fraction of calcium-polluted white dwarfs along the cooling sequence \citep{clsj24pca}.

Access to tens of thousands of white dwarfs with photometric and spectroscopic data has also led to advances in the automation of analysis processes, including the use of machine learning techniques. Notable examples in the white dwarf field include the use of neural networks for the spectral classification of $36\,500$ SDSS spectra \citep{vincent23,vincent25} or the classification using the coefficients of the BP/RP low-resolution spectra from {\it Gaia} DR3 \citep{garciazamora23,garciazamora25,vincent24}. Dimensionality reduction techniques have also been implemented to study {\it Gaia}'s BP/RP spectra \citep{kao24,perezcouto24,perezcouto25} and the medium-resolution spectra from DESI EDR \citep{byrne24}. In this context, machine learning tools allow the selection and characterization of outliers and anomalous objects in large databases. Examples include self-organizing maps in {\it Gaia} \citep{fustes13}, the analysis of SDSS spectra \citep{baron13,vincent25}, artificial neural networks directly applied to images \citep[e.g.,][]{margalef20,storey21,tanaka22}, automatic anomaly detection in time series \citep[e.g.,][]{malanchev21,muthukrishna22}, photometric redshifts \citep{singal22,dennis25} or radio observations \citep{mersarcik23}, and general methodologies applicable to any type of astronomical data (e.g., \texttt{ASTRONOMALY}; \citealt{astronomaly}).

This work presents the white dwarf catalog based on J-PLUS DR3, as a complement to the {\it Gaia} catalog by GF21. The catalog includes atmospheric parameters (effective temperature, surface gravity), composition (H-dominated, He-dominated, presence of polluting metals), and derived properties (mass, dust de-reddened $r-$band magnitude) for $14\,844$ white dwarfs, obtained with Bayesian SED-fitting techniques. The methodologies developed in previous studies \citep{clsj22pda, clsj24pca} are complemented by a novel analysis of outlier sources, defined as those largely incompatible with the theoretical models used in the fitting process. To achieve this, we applied the uniform manifold approximation and projection\footnote{\url{https://umap-learn.readthedocs.io}} ($\texttt{UMAP}$; \citealt{umap}) dimensionality reduction technique to the residuals of the best-fitting model across the $12$ J-PLUS bands, allowing the classification of outliers into different categories. The J-PLUS DR3 white dwarf catalog will be the starting point for future statistical analysis.

This paper is organized as follows. In Sect.~\ref{sec:data}, we describe the J-PLUS DR3 photometric data and the reference white dwarf catalog from GF21 used in our analysis. We present the methodology for obtaining atmospheric parameters, composition, and physical properties of the analyzed white dwarfs in Sect.~\ref{sec:method}. Sect.~\ref{sec:outliers} details the analysis of outliers using dimensionality reduction techniques. A comparison of the J-PLUS classification with previous studies is discussed in Sect.~\ref{sec:class}. Finally, the conclusions are provided in Sect.~\ref{sec:conclusion}. Magnitudes are expressed in the AB system \citep{oke83}.

\section{Data}\label{sec:data}

\begin{table} 
\caption{J-PLUS photometric system.}
\label{tab:filters}
\centering 
        \begin{tabular}{l c c c}
        \hline\hline\rule{0pt}{3ex} 
        Passband   & Effective wavelength   & FWHM  \\\rule{0pt}{2ex} 
                &   [nm]                & [nm]                  \\
        \hline\rule{0pt}{2ex}
        $u$             &353.3  &50.8            \\ 
        $J0378$         &378.2  &16.8            \\ 
        $J0395$         &393.9  &10.0            \\ 
        $J0410$         &410.8  &20.0            \\ 
        $J0430$         &430.3  &20.0            \\ 
        $g$             &479.0  &140.9           \\ 
        $J0515$         &514.1  &20.0            \\ 
        $r$             &625.7  &138.8           \\ 
        $J0660$         &660.4  &13.8            \\ 
        $i$             &765.6  &153.5           \\ 
        $J0861$         &861.0  &40.0            \\ 
        $z$             &896.5  &140.9           \\ 
        \hline 
\end{tabular}
\end{table}

\begin{figure*}[ht!]
\centering
\resizebox{0.49\hsize}{!}{\includegraphics{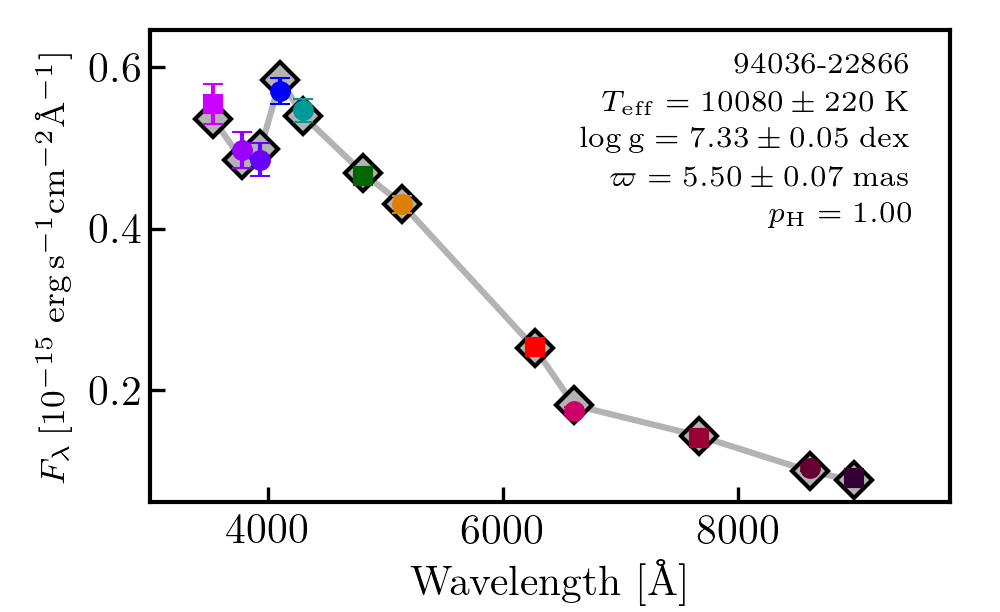}}
\resizebox{0.49\hsize}{!}{\includegraphics{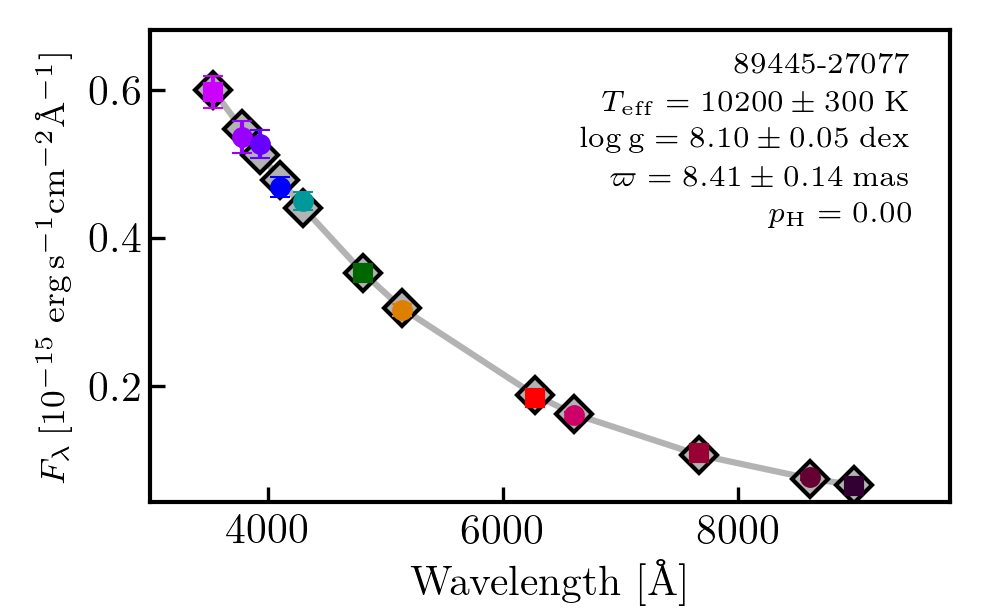}}
\resizebox{0.49\hsize}{!}{\includegraphics{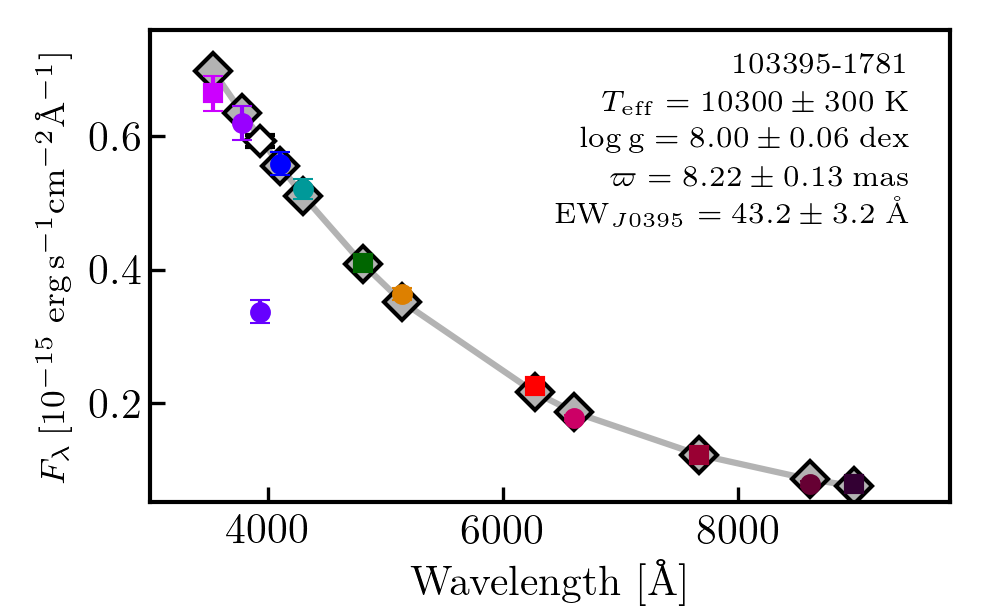}}
\resizebox{0.49\hsize}{!}{\includegraphics{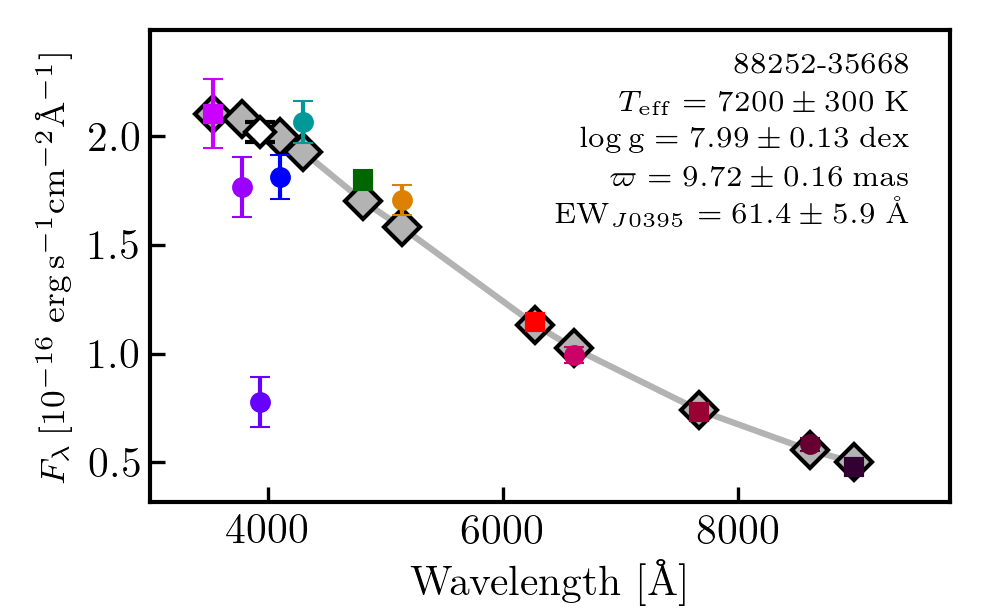}}
\caption{Spectral energy distributions of four white dwarfs analyzed in this work. The colored points in all panels represent the J-PLUS $3$-arcsec-diameter photometry, corrected for aperture effects. Broad-band filters ($u,g,r,i,z$) are shown as squares, whereas narrow- and medium-band filters ($J0378$, $J0395$, $J0410$, $J0430$, $J0515$, $J0660$, $J0861$) are shown as circles. The best-fitting solution is indicated gray diamonds connected with a solid line, with the derived parameters and their uncertainties labeled in the panels: effective temperature ($T_{\rm eff}$), surface gravity ($\log {\rm g}$), and parallax ($\varpi$). The unique J-PLUS identification, composed of the \texttt{TILE\_ID} of the reference $r-$band image and the \texttt{NUMBER} assigned by SExtractor to the source, is also reported in the panels for reference. {\it Top panels}: All passbands were used in the SED-fitting process, and the probability of having an H-dominated atmosphere ($p_{\rm H}$) is provided. {\it Bottom panels}: The equivalent width in the $J0395$ (${\rm EW}_{J0395}$) was computed to identify the presence of calcium absorption, using the expected continuum from the SED-fitting to the other eleven passbands (white diamond) as reference. These two sources have a probability of calcium absorption of $p_{\rm Ca} = 1$.}
\label{fig:examples}
\end{figure*}

\subsection{J-PLUS photometric data}\label{sec:jplus}
J-PLUS is being carried out at the Observatorio Astrof\'isico de Javalambre (OAJ, Teruel, Spain; \citealt{oaj}) using the 83 cm Javalambre Auxiliary Survey Telescope (JAST80) and T80Cam, a wide-field camera with a resolution of 9.2k $\times$ 9.2k pixels, offering a $2\deg^2$ field of view and a pixel scale of 0.55 arcsec pix$^{-1}$ \citep{t80cam}. The survey employs a system of $12$ optical passbands (Table~\ref{tab:filters}), designed for precise stellar classification and studies in the nearby Universe. Details on the observational strategy, data reduction, and scientific objectives of J-PLUS are described in \citet{cenarro19}.

The J-PLUS DR3 dataset includes $1\,642$ pointings, corresponding to $3\,284$ deg$^2$. The limiting magnitudes (5$\sigma$, 3 arcsec aperture) in DR3 reach around $22.5$ mag in the $g$ and $r$ bands and $21.5$ mag in the remaining ten filters. The median full width at half maximum (FWHM) of the point spread function (PSF) in the $r$ band is 1.1 arcsec. We performed source extraction using the $r$ band images with \texttt{SExtractor} \citep{sextractor}, and flux measurements in all $12$ bands were obtained at the predefined source positions based on the $r$ band detections. Regions affected by image borders, bright stars, or optical artifacts were masked from the initial area, yielding a final high-quality area of $2\,881$ deg$^2$. The J-PLUS DR3 is publicly accessible through the J-PLUS website\footnote{\url{www.j-plus.es/datareleases/data_release_dr3}}.

We performed aperture photometry using a $3$ arcsec diameter to study the white dwarf population. The measured fluxes were stored in the vector $\vec{f} = \{ f_j \}$, while their associated uncertainties were recorded in the vector $\sigma_{\vec{f}} = \{\sigma_j\}$, where the index $j$ corresponds to the J-PLUS passbands. The error estimates include contributions from photon noise, sky background fluctuations, and uncertainties in the photometric calibration \citep{clsj24gphot}.

\subsection{{\it Gaia} white dwarf catalog}\label{sec:gf21}
The {\it Gaia}-based catalog of white dwarfs presented in GF21 was used as a reference. The selection of the GF21 catalog can be summarized as follows: $1\,280\,266$ objects were selected using their location on the color absolute magnitude diagram of {\it Gaia} EDR3 and after applying quality flags. The SDSS DR16 \citep{sdss_dr16} spectroscopic information for $22\,998$ white dwarfs and $7\,124$ contaminants (main-sequence stars, subdwarfs, and quasars) was used to estimate the white dwarf probability, $P_{\rm WD}$, for each source in the initial sample. We refer the reader to GF21 for a detailed description of the selection criteria and the properties of the reference white dwarf sample.

The $359\,073$ sources with $P_{\rm WD} > 0.75$ in the GF21 catalog were cross-matched with J-PLUS DR3 using a $1.5$ arcsec radius. A total of $19\,581$ sources with $r \leq 20.5$ mag were found in common. The imposed magnitude limit ensures a well-defined volume and magnitude selection in the final white dwarf sample, as detailed in Sect.~\ref{sec:final}.

\subsection{Ancillary data}
At various stages, the visual and machine learning spectroscopic classification available at the Montreal White Dwarf Database\footnote{\url{http://www.montrealwhitedwarfdatabase.org}} \citep[MWDD;][]{MWDD} and the SIMBAD astronomical database\footnote{\url{http://simbad.u-strasbg.fr/simbad}} \citep{simbad} were used. Following the classification of \citet{sion83}, the main white dwarf spectral types considered in this work include those exhibiting hydrogen lines (DA), \ion{He}{II} lines (DO), \ion{He}{I} lines (DB), metal lines (DZ), carbon lines (DQ), featureless spectra (DC), magnetic fields (DH), and a main-sequence companion (WDMS).

\section{J-PLUS DR3 white dwarf catalog}\label{sec:method}
The methodology employed in the estimation of white dwarf atmospheric parameters and composition using a Bayesian framework is fully described in \citet{clsj22pda,clsj24pca}. The key aspects covered include the estimation of the likelihood and the assumed prior information, as summarized in the following sections. Representative examples of the analyzed white dwarfs are presented in Fig.~\ref{fig:examples}.

\begin{figure}[t]
\centering
\resizebox{\hsize}{!}{\includegraphics{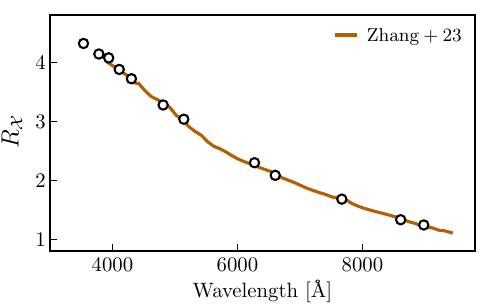}}
\caption{Extinction coefficients of the J-PLUS passbands scaled by $0.965$ (circles). The solid line depicts the extinction curve derived by \citet{zhang23} in the optical range.
}
\label{fig:eden_ext}
\end{figure}

\subsection{Likelihood}
The likelihood quantifies the probability of obtaining the fluxes observed in a $3$-arcsec aperture by J-PLUS under the assumption of a particular model. The absolute model fluxes were derived from theoretical white dwarf atmosphere models, which depend on the effective temperature ($\teff$) and surface gravity ($\log {\rm g}$). These absolute model fluxes were converted to apparent fluxes with the parallax, which was used as an additional free parameter in the fitting process, the interstellar extinction, and the aperture corrections that translate total fluxes to $3$-arcsec apertures\footnote{The aperture corrections can be derived from the table \texttt{jplus.FNuDualPointSources} in the J-PLUS database, and their estimation is detailed in \citet{clsj22pda}}. The theoretical fluxes were then compared with the observed photometric measurements in the $12$ J-PLUS passbands, weighted by their Gaussian uncertainties.

The extinction correction was updated with respect to \citet{clsj22pda} and is based on the 3D extinction map of \citet{edenhofer24}, which builds on the stellar distances and extinctions provided by \citet{zhang23}. A Bayesian approach was employed, modeling interstellar extinction for $54$ million nearby stars as a function of distance and incorporating a Gaussian-process prior methodology to mitigate fingers-of-God artifacts. We obtained the integrated median value of the extinction parameter $E$ at white dwarf location for each parallax (i.e., distance) using the \texttt{dustmaps}\footnote{\url{https://github.com/gregreen/dustmaps}} program \citep{dustmaps}. The extinction coefficient\footnote{Accessible in the table \texttt{jplus.Filter} of the J-PLUS database.} for filter $\mathcal{X}$, denoted $R_{\mathcal{X}}$, was estimated with the star-pair technique described in \citet{yuan13} applied to J-PLUS DR1 data. The J-PLUS extinction coefficients were found to be compatible with the extinction curve derived by \citet{zhang23} after applying a factor $0.965$ (Fig.~\ref{fig:eden_ext}). Therefore, the extinction for each white dwarf in the fitting process was estimated as $0.965\,R_{\mathcal{X}} \times E\,(\varpi)$. Finally, the 3D map provided by \citet{edenhofer24} covers the $69 - 1\,250$ pc range. The integrated extinction $E$ at $d < 69$ pc was considered equal to zero, and our final sample is limited to a distance of $1$~kpc (Sect.~\ref{sec:final}).

We investigated two atmospheric compositions, denoted as $t$, corresponding to H- and He-dominated atmospheres. Hydrogen-dominated atmospheres were modeled using pure-H configurations ($t = {\rm H}$, \citealt{tremblay11,tremblay13}). In contrast, He-dominated atmospheres were described with mixed models, assuming an H/He ratio of $10^{-5}$ for $T_{\rm eff} > 6\,500$ K, and pure-He models for $T_{\rm eff} < 6\,500$ K ($t = {\rm He}$, \citealt{cukanovaite18, cukanovaite19}). The use of mixed models for He-dominated atmospheres is motivated by observational evidence indicating the need for an additional electron donor in helium atmospheres to reconcile the mass distributions of H- and He-dominated white dwarfs \citep{bergeron19}. From a modeling perspective, adopting an H/He ratio of $10^{-5}$ serves as a suitable ansatz. The leading physical candidate for the additional donor is dredged-up carbon, although its abundance is expected to be too low to produce detectable features in optical spectra \citep{camisassa23,blouin23,blouin23uv}. The mass-radius relation from \citet{fontaine01} was adopted, considering thin hydrogen layers for helium atmospheres and thick hydrogen layers for H-dominated atmospheres. A detailed discussion of these models can be found in, e.g., \citet[][]{bergeron19,GF20,40pcii}.

\begin{figure}[t]
\centering
\resizebox{\hsize}{!}{\includegraphics{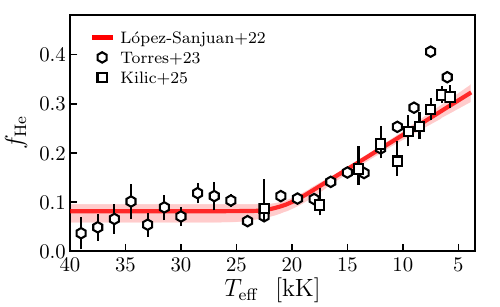}}
\caption{Fraction of He-dominated white dwarfs ($f_{\rm He}$) as a function of effective temperature ($T_{\rm eff}$). The solid red line represents the result from \citet{clsj22pda} and corresponds to the prior used in Sect.~\ref{sec:prior}. Measurements from \citet{torres23}, using {\it Gaia} BP/RP spectra (hexagons), and from \citet{kilic25}, using the $100$ pc spectroscopic sample (squares), are also shown.}
\label{fig:fhe}
\end{figure}

\subsection{Priors}\label{sec:prior}
We assumed a prior in the parallax: 
\begin{equation}
        P\,(\varpi) = P_G (\varpi\,|\,\varpi_{\text{EDR3}}, \sigma_{\varpi}),
\end{equation}
where $\varpi_{\text{EDR3}}$ is the {\it Gaia} EDR3 parallax \citep{gaiaedr3} corrected by systematic offsets following \citet{lindegren21a}, $\sigma_{\varpi}$ is the error in the parallax, and $P_{\rm G}$ depicts a Gaussian probability function.

\begin{figure*}[t]
\centering
\resizebox{\hsize}{!}{\includegraphics{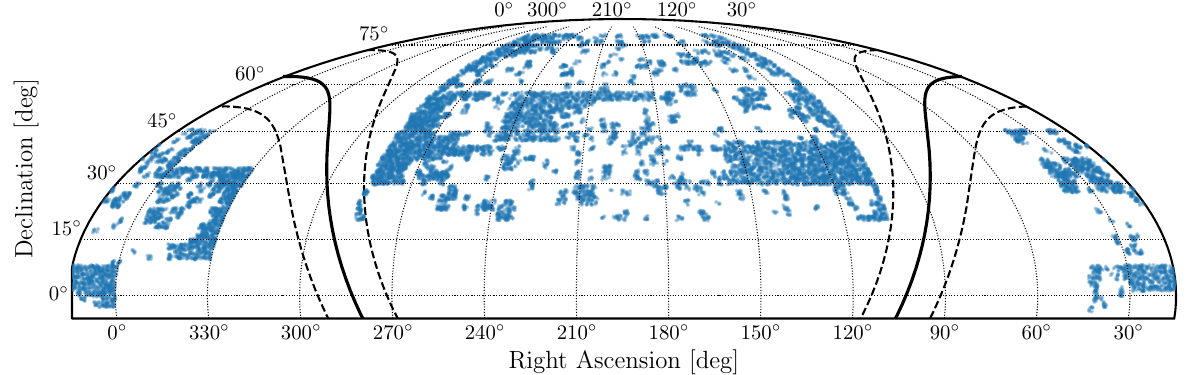}}
\caption{Sky distribution of the $\nwd$ white dwarfs in the J-PLUS DR3 catalog. The solid line depicts the position of the Milky Way disk, and the dashed lines correspond to +/- $10$ deg in Galactic latitude.
}
\label{fig:wdsky}
\end{figure*}

\begin{figure}[t]
\centering
\resizebox{\hsize}{!}{\includegraphics{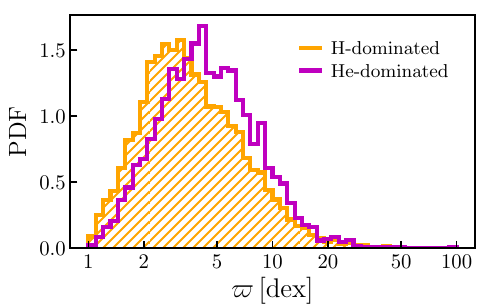}}
\caption{Probability distribution function of the parallax ($\varpi$) for H-dominated (hatched, orange) and He-dominated (solid, purple) atmospheres in the final J-PLUS DR3 catalog.
}
\label{fig:wddist}
\end{figure}

A prior in the atmospheric composition was also applied, providing the probability of a white dwarf having an H-dominated atmosphere at a given $\teff$ as
\begin{equation}
    P\,({\rm H}\,|\,T_{\text{eff}}) = 1 - f_{\rm He}\,(\teff),
\end{equation}
where $f_{\rm He}\,(\teff)$ is the fraction of He-dominated white dwarfs derived by \citet{clsj22pda} using J-PLUS DR2,
\begin{equation}
    f_{\rm He}\,(\teff) = 0.24 - 0.14\times \bigg( \frac{T_{\rm eff}}{10^4\ {\rm K}} - 1 \bigg),
\end{equation}
imposing a minimum value of $0.08$. The spectral evolution from \citet{clsj22pda} is in good agreement with previous estimations; however, new measurements have become available since the publication of their results (see \citealt{bedard24} for a review). For example, the comparison with \citet{torres23}, based on {\it Gaia} DR3 BP/RP spectra, and with \citet{kilic25}, based on the $100$ pc spectroscopic sample, is presented in Fig.~\ref{fig:fhe}. A remarkable agreement is found, reinforcing the results from J-PLUS DR2 and further justifying the assumed prior.

\subsection{Posterior parameters and derived quantities}\label{sec:posterior}
We combined the likelihood and the priors using Bayes' theorem to obtain the posterior distribution functions (PDFs) of the parameters $\theta = \{ \teff, \logg, \varpi \}$ for two compositions $t = \{ {\rm H, He} \}$. 

The point estimates of the parameters and their uncertainties were obtained with a Gaussian fit to the PDFs. The probability of having an H-dominated atmosphere was estimated as
\begin{equation}
    p_{\rm H} = \int {\rm PDF}\,({\rm H},\theta)\,{\rm d}\theta,
\end{equation}
and the probability of being He-dominated is $p_{\rm He} = 1 - p_{\rm H}$. Derived quantities, such as the stellar mass ($M$), the total $r$ band apparent magnitude corrected for dust extinction ($\hat{r}$), or the effective volume accessible for each white dwarf given the selection of the sample ($V_{\rm eff}$), where estimated as the PDF-weighted solutions from the models.

The probability of presenting calcium absorption ($\pca$) was also estimated as an additional diagnostic, independent of the H- or He-dominated atmosphere classification, following \citet{clsj24pca}. In this case, we applied the same Bayesian SED fit process, but the $J0395$ passband, which is sensitive to calcium absorption, was excluded from the analysis. The modeling provided the continuum level in the absence of polluting metals, which we then compared with the observed flux to measure the equivalent width ${\rm EW}_{J0395}$ and its uncertainty. The equivalent-width distribution was modeled to obtain the fraction of calcium white dwarfs as a function of $\teff$,
\begin{equation}
    f_{\rm Ca} = 0.067 - 0.18\times \bigg( \frac{T_{\rm eff}}{10^4\ {\rm K}} - 1 \bigg),
\end{equation}
imposing a minimum value of $0.01$ and a maximum of 1. We used this evolution as a prior to properly weight the significance of ${\rm EW}_{J0395}$ and provide $\pca$. We refer the reader to \citet{clsj24pca} for an extensive discussion, including the selection and confirmation of new metal-polluted white dwarfs.

\begin{figure}[t]
\centering
\resizebox{\hsize}{!}{\includegraphics{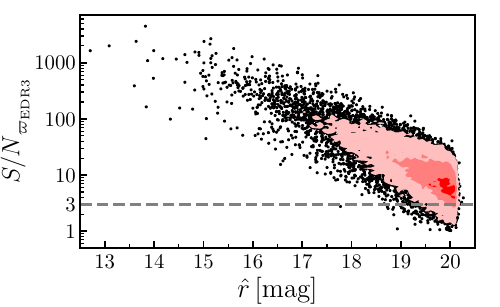}}
\caption{Signal-to-noise ratio in the {\it Gaia} EDR3 parallax (${\rm SNR}_{\varpi_{\rm EDR3}}$) used as a prior versus the de-reddened$r$ band apparent magnitude ($\hat{r}$) for the white dwarf sample. The black dots show individual measurements. The red areas, from lighter to darker, enclose 90\%, 50\%, and 10\% of the sources, respectively. The dashed gray line marks ${\rm SNR}_{\varpi_{\rm EDR3}} = 3$.}
\label{fig:snrpx}
\end{figure}

\subsection{Final sample}\label{sec:final}
We analyzed the $19\,581$ sources with $r \leq 20.5$ mag in common with the GF21 white dwarf catalog following the methodology described in previous sections. A selection with $\hat{r} < 20$ mag and $1 \leq \varpi < 100$ mas was then applied to the posterior in the variables, ensuring a well-defined selection function, and only sources with a selection probability $p_{\rm sel} \geq 0.01$ within these limits were kept in the final catalog. This reduced the sample to $\nwd$ white dwarfs, corresponding to a $p_{\rm sel}$-weighted number of $13\,070.86$ sources and an effective density of $4.5$ source deg$^{-2}$ (Fig.~\ref{fig:wdsky}). The final catalog, available at the J-PLUS database and the Centre de Donn\'ees astronomiques de Strasbourg (CDS), is described in Appendix~\ref{app:catalog}. 

The parallax distribution of the final sample is shown in Fig.~\ref{fig:wddist}. The distribution for H-dominated white dwarfs peaks at $\varpi \approx 3$ mas, while for He-dominated white dwarfs it peaks at $\varpi \approx 4.5$ mas. This difference is caused by the combination of the physical properties of the sample with the selection $\hat{r}<20$ mag: at small parallaxes, only the hottest white dwarfs are detected, where the fraction of He-dominated atmospheres tends to be $5-10$\% (Fig.~\ref{fig:fhe}), while at larger parallaxes, cooler white dwarfs are included in the sample and $f_{\rm He}$ tends to $25$\%. As a result, a peak at larger parallax is effectively produced for He-dominated atmospheres compared to H-dominated ones. It should be recalled that the variable $V_{\rm eff}$ in the catalog provides the effective detection volume for each white dwarf, allowing selection effects to be corrected. Regarding the quality of the prior parallax information from {\it Gaia} EDR3 used in the fitting process, it was found that 96\% of the sources have a signal-to-noise ratio ($S/N$) in the parallax measurement of $S/N > 3$, with $85$\% having $S/N > 5$ (Fig.~\ref{fig:snrpx}). 

\begin{figure}[t]
\centering
\resizebox{\hsize}{!}{\includegraphics{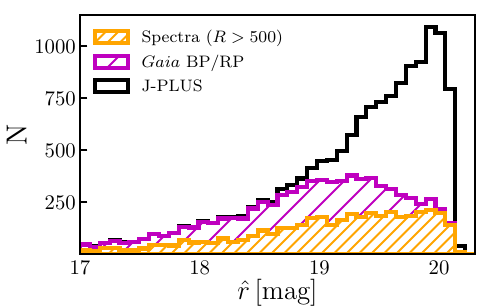}}
\caption{Stacked histogram in $\hat{r}$-band magnitude showing the available atmospheric composition: spectroscopic with $R \gtrsim 500$ (hatched, orange), {\it Gaia} BP/RP spectra (hatched, purple), and J-PLUS (solid, black).}
\label{fig:spec}
\end{figure}

Spectroscopic classification is available for $56$\% of the sources in the catalog, corresponding to $4\,407$ ($29$\%) with $R \gtrsim 500$ (SDSS DR17, \citealt{vincent23}; DESI EDR, \citealt{desi_edr_wd}; 40 pc sample, \citealt{40pciv}; 100 pc sample, \citealt{kilic25}), and $3\,969$ ($27$\%) from {\it Gaia} low-resolution BP/RP spectra \citep{vincent24} without higher-resolution information. Consequently, $6\,468$ ($44$\%) white dwarfs in the catalog are classified solely on the basis of J-PLUS photometry. The magnitude distribution of these subsamples is presented in Fig.~\ref{fig:spec}. Sources with spectra dominate at $\hat{r} < 19$ mag, with only $6$\% lacking spectroscopic classification. At fainter magnitudes, J-PLUS becomes the primary resource, with $60$\% of the sources lacking spectroscopic data. While ongoing and future spectroscopic surveys are expected to increase the number of classified objects, J-PLUS will continue to provide valuable photometric information, even with full spectroscopic coverage.

The validation of the $\teff$ and $\logg$ measurements from J-PLUS photometry is presented in \citet{clsj22pda,clsj24pca}. In summary, the typical uncertainty in $\teff$ at temperatures lower than $30\,000$~K is $5$\%, while the uncertainty in $\logg$ is mainly dictated by the {\it Gaia} parallax and is $0.15$ dex at $\teff < 30\,000$~K. The J-PLUS spectral classification is further tested in Sec.~\ref{sec:class} using those sources with spectroscopic information.

\begin{figure*}[t]
\centering
\resizebox{0.49\hsize}{!}{\includegraphics{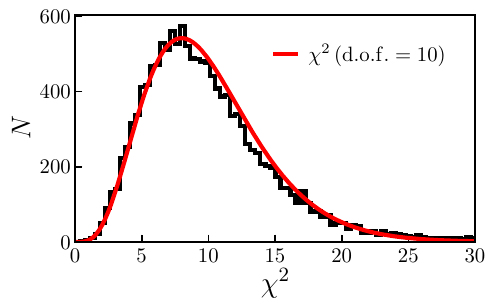}}
\resizebox{0.49\hsize}{!}{\includegraphics{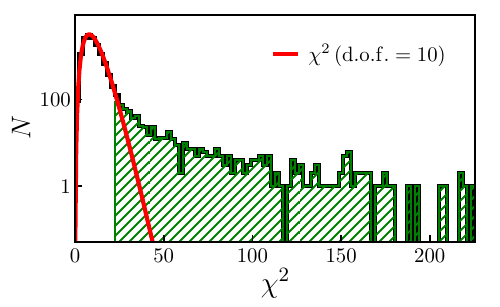}}
\caption{Distribution of $\chi^2$ values for the white dwarf sample (black histograms). The red lines represent the theoretical $\chi^2$ distribution with $10$ degrees of freedom. The right panel shows the same data with the $y-$axis in logarithmic scale and an expanded $x-$axis range. The green histogram highlights the $771$ outlier sources with $\chi^2 \geq 23.2$.
}
\label{fig:chidist}
\end{figure*}

\section{Dimensionality reduction to investigate outlier white dwarfs}~\label{sec:outliers}
This section is devoted to the analysis of the outlier population, as defined in Sect.~\ref{sec:out}, using dimensionality reduction tools to explore the $12$-dimensional J-PLUS data in a two-dimensional space (Sect.~\ref{sec:umap}). The physical origin of the different populations found is explored in Sect.~\ref{sec:pop}.

\subsection{Definition of outliers}\label{sec:out}
We defined the $\chi^2$ parameter as
\begin{equation}
    \chi^2 = \sum_j \frac{(f_j - f^{\rm best}_j)^2}{\sigma^2_{j}} = \sum_j  \frac{(\vec{f} - \vec{f}^{\rm best})^2}{\sigma^2_{\vec{f}}},
\end{equation}
where the index $j$ runs the $12$ J-PLUS passbands, and $\vec{f}^{\rm best} = \{f^{\rm best}_j\}$ is the best-fitting model vector following the procedure described in Sect.~\ref{sec:posterior} to estimate $p_{\rm Ca}$. This minimizes the impact of metal pollution in the SED-fitting process.

\begin{figure*}[t]
\centering
\resizebox{0.49\hsize}{!}{\includegraphics{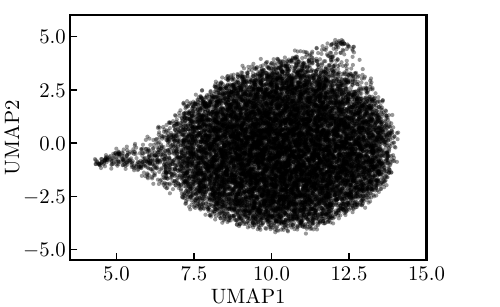}}
\resizebox{0.49\hsize}{!}{\includegraphics{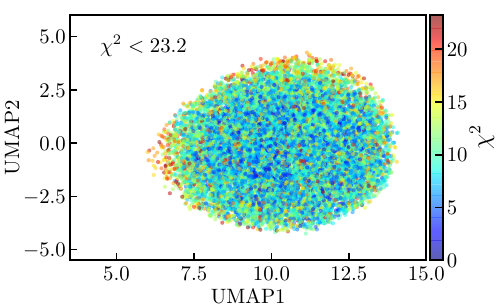}}
\resizebox{0.49\hsize}{!}{\includegraphics{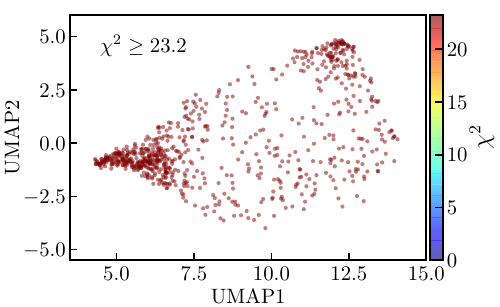}}
\resizebox{0.49\hsize}{!}{\includegraphics{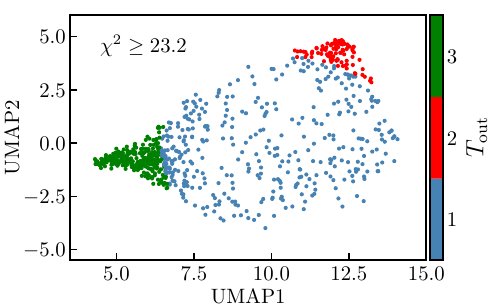}}
\caption{Two-dimensional $\texttt{UMAP}$ manifold of the J-PLUS 12-dimensional residuals space, $(\vec{f} - \vec{f}^{\rm best})/\sigma_{\vec{f}}$. {\it Top-left panel}: Distribution of the full sample (black dots). {\it Top-right panel}: Distribution of the sources with $\chi^2 < 23.2$, as shown in the color scale. {\it Bottom-left panel}: Distribution of the outliers with $\chi^2 \geq 23.2$. {\it Bottom-right panel}: Definition of the three types of outliers ($\tout$), as indicated in the color bar.
}
\label{fig:umap}
\end{figure*}

The $\chi^2$ distribution of the $\nwd$ white dwarfs in the catalog is presented in the left panel of Fig.~\ref{fig:chidist}. The residuals are found to follow a $\chi^2$ distribution with $10$ degrees of freedom (d.o.f), which is consistent with the $12$ photometric points from J-PLUS and the two parameters in the fit, corresponding to the effective temperature and spectral type \citep{clsj22pda}. Surface gravity is not considered an effective parameter because it is well constrained by the {\it Gaia} parallax prior, coupled with the assumption of a mass-radius relation in the models.

When the high-$\chi^2$ tail is analyzed (right panel in Fig.~\ref{fig:chidist}), a clear excess of objects is found. Taking $\chi^2 \geq 23.2$ as a reference, corresponding to a 99\% cumulative probability in the expected distribution, 771 sources (5\% of the sample) are observed, whereas only $148$ would be expected. Hereafter,  sources with $\chi^2 \geq 23.2$ were defined as outliers and those with $\chi^2 < 23.2$ as inliers. This approach is similar to that presented in \citet{covey07} to spot outlier stars in a seven-dimensional color space.

\subsection{Analysis of outlier white dwarfs using $\texttt{UMAP}$ }\label{sec:umap}
In the present study, we used $\texttt{UMAP}$ to analyze the white dwarf outliers. This dimensionality reduction technique is primarily used for visualizing high-dimensional data in a lower-dimensional space, typically 2D. It is based on manifold learning and topological data analysis, aiming to preserve both the local and global structure of the data. The $\texttt{UMAP}$ algorithm works by constructing a graph of the data neighbors and optimizing a layout in the lower-dimensional space to best maintain the relationships between points.

We applied the $\texttt{UMAP}$ algorithm to the normalized residuals $(\vec{f} - \vec{f}^{\rm best})/\sigma_{\vec{f}}$ instead of to the J-PLUS photometric data. This approach has three advantages. First, the twelve input variables are naturally normalized to a common meaningful scale. Second, it reduces most of the population to random noise, eliminating the structure present in the fluxes due to variations in effective temperature, surface gravity, and main composition. Finally, the random population allows us to define an area in the $\texttt{UMAP}$ manifold consistent with statistical fluctuations, amplifying the presence of physical populations that exhibit a systematic pattern in the residuals. This last point should enable the identification of white dwarfs not represented by the theoretical models used, such as DZs, DQs, or binary systems.

The algorithm has two parameters, $\texttt{n\_neighbors}$ and $\texttt{min\_dist}$. The $\texttt{n\_neighbors}$ parameter controls the number of neighboring points considered when constructing the local neighborhood graph. A smaller value tends to preserve local structure, while a larger value captures more global relationships. The $\texttt{min\_dist}$ parameter determines the minimum distance between points in the embedded space, influencing the clustering of points. Smaller values allow for denser clusters, while larger values lead to more spread-out embeddings. Several values for both parameters were explored, and the overall structure of the 2D manifold was found to be preserved regardless of the chosen parameters. We set the final configuration to $\texttt{n\_neighbors}=100$ and $\texttt{min\_dist}=0.5$. To ensure reproducibility, $\texttt{random\_state}=1906$ was used. The 2D $\texttt{UMAP}$ manifold of the J-PLUS 12-dimensional residuals is shown in the top-left panel of Fig.~\ref{fig:umap}. Three main populations were found: a circular concentration plus two appendices, one at UMAP1 $< 7.5$ and another at UMAP2 $> 2.5$. In These populations are analyzed in the next section.

\begin{figure}[t]
\centering
\resizebox{\hsize}{!}{\includegraphics{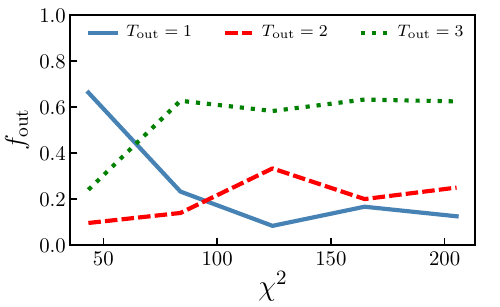}}
\caption{Relative fraction of each outlier type, $f_{\rm out}$, as a function of $\chi^2$. The blue, green, and red lines depict $f_{\rm out}$ for $\tout = 1, 2$, and $3$, respectively.
}
\label{fig:fout}
\end{figure}

\subsection{Understanding the $\texttt{UMAP}$ distribution}\label{sec:pop}
To understand the three populations in the $\texttt{UMAP}$ space, the locations of outliers and inliers were first explored. The inliers with $\chi^2 < 23.2$ are concentrated in the main circular population, exhibiting a trend for larger $\chi^2$ near the borders (top-right panel in Fig.~\ref{fig:umap}). Outliers with $\chi^2 \geq 23.2$ populate the two appendices (bottom-left panel in Fig.~\ref{fig:umap}). Notably, some outliers are also found in the main circular region.

The region populated by the inliers defines the location of random noise sources, a fact that was used to select outliers compatible with random fluctuations in the photometry. We performed a novelty detection using the Local Outlier Factor ($\texttt{LOF}$) function from $\texttt{scikit-learn}$\footnote{\url{https://scikit-learn.org}} trained on sources with $\chi^2 < 16.0$. This training set encompasses $90$\% of the $\chi^2$ distribution and ensures a pure random sample. The parameters used where $\texttt{n\_neighbors} = 250$, $\texttt{contamination} = 0.001$, and $\texttt{novelty} = \texttt{True}$. The model was then applied to the outlier sources, and an outlier type $T_{\rm out}$ was assigned as follows (bottom-right panel in Fig.~\ref{fig:umap}): $\tout = 1$ for white dwarfs classified as inlier by \texttt{LOF} ($391$ sources), $\tout = 2$ for \texttt{LOF} outliers with UMAP1 $> 9$ ($98$ sources), and $\tout = 3$ for \texttt{LOF} outliers with UMAP1 $< 9$ ($282$ sources). We assigned $\tout = 0$ to the remaining $14\,073$ white dwarfs with $\chi^2 < 23.2$. 

\begin{figure}[t]
\centering
\resizebox{\hsize}{!}{\includegraphics{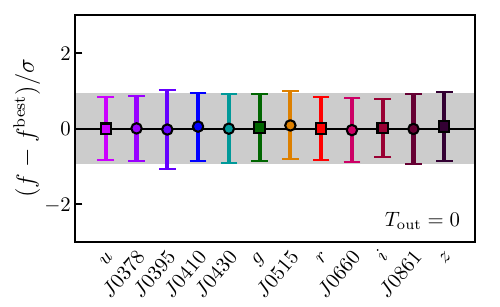}}
\resizebox{\hsize}{!}{\includegraphics{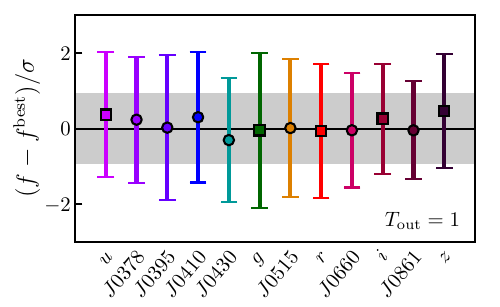}}
\caption{Normalized residuals for the different J-PLUS photometric bands, as labeled on the $x-$axis. The median and the dispersion of the best Gaussian describing the residuals of the population are shown with the symbols and the error bars, respectively. The solid black line depicts a zero difference, and the shaded region represents the expected range for a random distribution. The top panel corresponds to the outlier type $\tout = 0$ and the bottom panel to $\tout = 1$. 
}
\label{fig:chi_tout01}
\end{figure}

The relative fractional contribution of each outlier type to the total outlier population, $f_{\rm out}$, and its variation with $\chi^2$ is presented in Fig.~\ref{fig:fout}. We find that $\tout = 1$ dominates at the lower $\chi^2$, with $f_{\rm out} \approx 0.65$, and then steadily declines to $f_{\rm out} \approx 0.10$ above $\chi^2 = 100$. The situation is the opposite for the other two outlier types, which increase in relative importance and reach $f_{\rm out} \approx 0.20$ for $\tout = 2$ and $f_{\rm out} \approx 0.65$ for $\tout = 3$ at $\chi^2 > 100$. In the following sections, we examine each type of outlier to better understand their nature.

\begin{figure}[t]
\centering
\resizebox{\hsize}{!}{\includegraphics{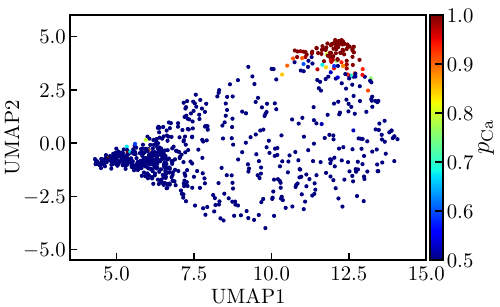}}
\resizebox{\hsize}{!}{\includegraphics{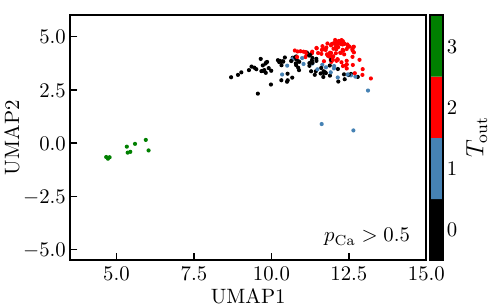}}
\caption{Two-dimensional $\texttt{UMAP}$ manifold for the J-PLUS 12-dimensional residuals space. {\it Top panel}: Distribution of the outliers color-coded by their $\pca$. {\it Bottom panel}: Distribution of all sources with $\pca > 0.5$ color-coded with $\tout$.
}
\label{fig:umap_pca}
\end{figure}

\begin{figure}[t]
\centering
\resizebox{\hsize}{!}{\includegraphics{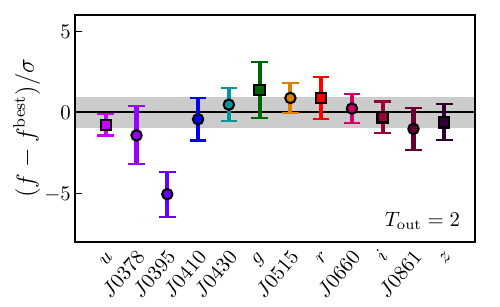}}
\resizebox{\hsize}{!}{\includegraphics{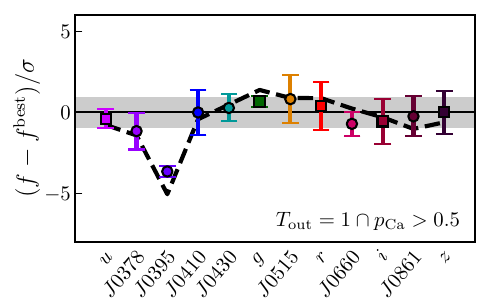}}
\resizebox{\hsize}{!}{\includegraphics{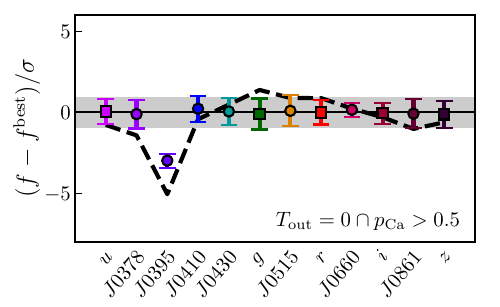}}
\caption{Same as shown in Fig.~\ref{fig:chi_tout01}, but for white dwarfs with $\tout = 2$ ({\it top panel}), $\tout = 1$ and $\pca > 0.5$ ({\it middle panel}), and $\tout = 0$ and $\pca > 0.5$ ({\it bottom panel}). The dashed lines depicts the median points in the $\tout = 2$ case for reference.
}
\label{fig:chi_tout2}
\end{figure}

\subsubsection{The random outliers}
To explore patterns in outlier populations, a Gaussian was fit to the distribution of $(f_j - f_j^{\rm best})/\sigma_j$ values. The obtained medians and dispersions for $\tout = 0$ and $\tout = 1$ sources are presented in Fig.~\ref{fig:chi_tout01}. The inliers present the expected behavior of random fluctuations: the obtained medians are close to zero and the dispersion is nearly $0.90$, similar to the expected $0.91$ for $12$ filters and $10$ d.o.f. \citep{clsj22pda}. Interestingly, white dwarfs with $\tout = 1$ also have a median close to zero but with enhanced dispersions, reaching $1.6$ on average. This suggests that the origin of the large $\chi^2$ may be an underestimation of the uncertainties (a plausible situation at the brighter magnitudes), inaccurate measurements in a limited number of bands, or small deviations between the assumed models and the actual composition, such as white dwarfs with mixed spectral types (DBA, DAB) or hot DOs (our He-dominated models only reach $T_{\rm eff} = 40\,000$~K, and above this effective temperature only H-dominated models are used). The higher fraction of $\tout = 1$ sources at low $\chi^2$ values with respect to $\tout = 2$ and $\tout = 3$ classes (Fig.~\ref{fig:fout}) is also consistent with a random nature, since $148$ sources within the outlier population are expected from the $\chi^2$ distribution properties and should concentrate near the $\chi^2 = 23.2$ boundary (Sect.~\ref{sec:out}).

In addition, ZZ Ceti variable white dwarfs were tested as a possible source of random outliers. Sources within the instability strip, as defined in \citet{tremblay15}, and with masses $M \geq 0.45\ M_{\odot}$ were selected. This selection resulted in a subset of $1\,783$ white dwarfs from our sample. Only $53$ ($3$\%) were classified as $\tout = 1$, with $7$ and $5$ classified as $\tout = 2$ and $\tout = 3$, respectively. The sources in the $\tout = 0$ category did not display any evident trend beyond randomness at any magnitude, with a median of zero and the expected dispersion in the normalized residuals. Therefore, we concluded that the J-PLUS observational strategy, which includes three consecutive exposures per passband and a typical time span of $45-90$ min to observe the $12$ filters, combined with the typical ZZ Ceti periods ($3-15$ min; \citealt{vincent20}) and amplitudes ($\lesssim 0.2$ mag; \citealt{vincent20}), makes the final J-PLUS photometry largely insensitive to their flux variability.

In summary, no clear physical population is found to be associated with the $T_{\rm out} = 1$ outliers, which appear to comprise a variety of origins. The large $\chi^2$ may arise from instrumental causes, such as underestimated errors or inaccurate measurements, or from physical causes; however, this does not result in a clear pattern in the residuals from the SED-fitting process. 

\subsubsection{The metal-polluted outliers}
The $\pca$ values for the outlier populations are presented in Fig.~\ref{fig:umap_pca}. White dwarfs with $\tout = 2$ are found to have a high probability of exhibiting calcium absorption and therefore can be assigned to the class of metal-polluted white dwarfs. The top panel of Fig.~\ref{fig:chi_tout2} shows the distribution of the normalized residuals. A large, $5.1\sigma$ flux deficit is observed in the $J0395$ band, and in the $u$ and $J0378$ passbands (although less significant). Conversely, a flux excess is present in the $g$, $J0515$, and $r$ passbands. This behavior is apparent in the source presented in the bottom-right panel in Fig.~\ref{fig:examples}.

\begin{figure}[t]
\centering
\resizebox{\hsize}{!}{\includegraphics{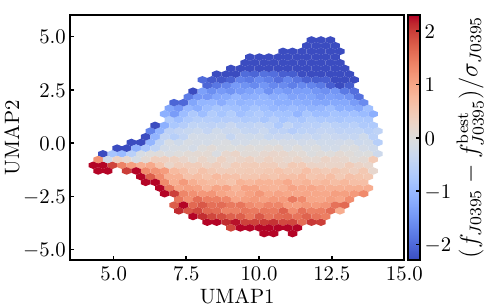}}
\caption{Median value of the normalized residuals in the $J0395$ passband along the \texttt{UMAP} two-dimensional manifold.
}
\label{fig:chi_umap_j0395}
\end{figure}

Despite the strong correlation between $\tout = 2$ outliers and large $\pca$ values, exceptions are found. First, some sources with $\pca > 0.5$ in the random-noise area (bottom panel in Fig.~\ref{fig:umap_pca}), clustered near the $\tout = 2$ region at UMAP2 $> 2.5$. We find $20$ with $\tout = 1$ and $75$ with $\tout = 0$, presenting a noticeable deficit in $J0395$ at the $3.6\sigma$ and $3.0\sigma$ levels, respectively, and approaching a zero difference in the other passbands (middle and bottom panels in Fig.~\ref{fig:chi_tout2}). There are also nine sources with $\pca > 0.5$ in the $\tout = 3$ population, but they can be discarded as false positives, as demonstrated in the following section. The $\teff$ distribution for $\pca > 0.5$ sources and different outlier types are similar, suggesting that the sequence represents an increase in calcium abundance from $\tout = 0$ to $\tout = 2$.

Only six sources have $\tout = 2$ and $\pca < 0.5$. These comprise three sources classified as metal-polluted, two magnetic white dwarfs, and one DC. This strengthens the capability of the $\tout = 2$ selection to identify metal-polluted systems.

Finally, motivated by the clustering of sources with $\pca > 0.5$ in the region where UMAP2 $> 2.5$, $575$ sources were selected with $\tout = 0$, $\pca < 0.1$, and UMAP2 $> 2.5$. A deficit in the $J0395$ filter is observed at a $1.8\sigma$ level. However, this signal can be ruled out as representative by performing the same analysis for the $549$ sources with $\tout = 0$, $\pca < 0.1$, and UMAP2 $< -3.2$, i.e., at the opposite end of the 2D manifold. In this case, an excess in the $J0395$ filter is obtained, with the same statistical significance of $1.8\sigma$. This suggests that the UMAP2 axis is dominated by the residual in the $J0395$  filter, a hypothesis confirmed by the analysis of the median residual along the \texttt{UMAP} manifold\footnote{For completeness, the median residual in the other J-PLUS passbands are presented in Appendix~\ref{app:chi_umap}} (Fig.~\ref{fig:chi_umap_j0395}). This highlights that the statistical approach used in the estimation of $\pca$ is critical, as spurious detections due to statistical fluctuations are minimized.

In summary, $\tout = 2$ is sensitive to the presence of different metals and can be used to select a high-confidence sample of metal-polluted white dwarfs, complementing the $\pca$ measurement based solely on the $J0395$ passband and sensitive only to Ca absorption.

\subsubsection{The two-component outliers}
\begin{figure}[t]
\centering
\resizebox{\hsize}{!}{\includegraphics{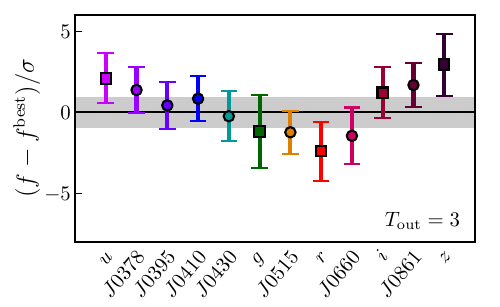}}
\caption{Same as shown in Fig.~\ref{fig:chi_tout01}, but for white dwarfs with $\tout = 3$.
}
\label{fig:chi_tout03}
\end{figure}

Analysis of the $\tout = 3$ sources reveals that most correspond to two-component systems consisting of a blue white dwarf and a redder component, but other types of white dwarf are also included, as analyzed in the latter part of this section. The red component is not always physically associated; in several cases, it results from contamination by a nearby red star or by a galaxy. The median residuals of the population with $\tout=3$ are shown in Fig.~\ref{fig:chi_tout03}. An excess is observed in both the bluer and the redder filters, while a deficit is present in the intermediate filters, mainly $J0515$, $r$, and $J0660$.

\begin{figure}[t]
\centering
\resizebox{\hsize}{!}{\includegraphics{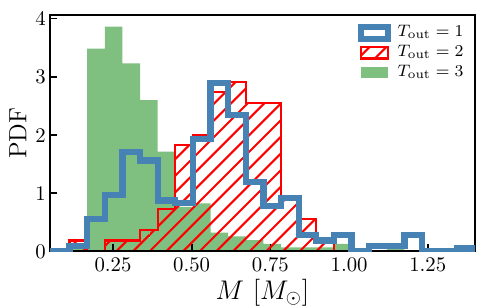}}
\caption{Probability distribution function of the estimated stellar mass for sources with $\tout = 1$ (empty histogram), $\tout = 2$ (hatched histogram), and $\tout = 3$ (filled histogram).
}
\label{fig:mass_tout}
\end{figure}

The inferred mass distribution of the outlier populations is presented in Fig.~\ref{fig:mass_tout}, with the mass estimated as $M_{\rm H} \times p_{\rm H} + M_{\rm He} \times (1-p_{\rm H})$. Extremely low-mass objects dominate the $\tout = 3$ population. This finding is consistent with the two-component scenario, as fitting these systems under the assumption of a single white dwarf often leads to an apparent over-luminous, low-mass solution. The distribution of $\tout = 1$ sources exhibits a population of low-mass objects, though peaking at higher masses than those in the $\tout = 3$ category. It should be noted that only systems displaying a significant deviation from random fluctuations are classified as $\tout = 3$, while additional two-component candidates, including double degenerates, can be identified based on their apparent low-mass solutions. Indeed, the fraction of sources with $M \leq 0.45$~$M_{\odot}$ is $0.17$, $0.32$, $0.10$, and $0.84$ when moving from $\tout = 0$ to $\tout = 3$.

\begin{figure}[t]
\centering
\resizebox{\hsize}{!}{\includegraphics{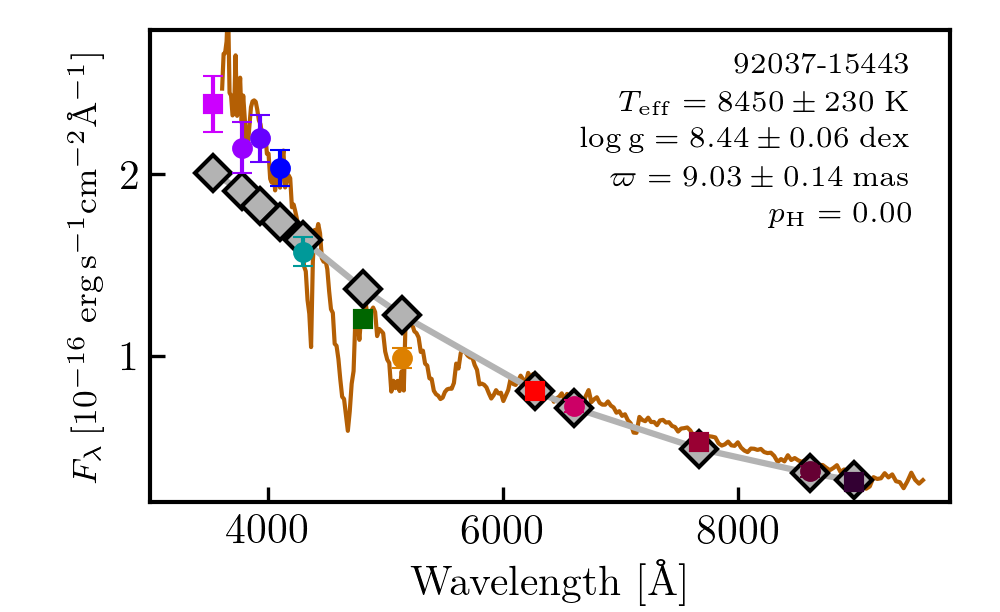}}
\resizebox{\hsize}{!}{\includegraphics{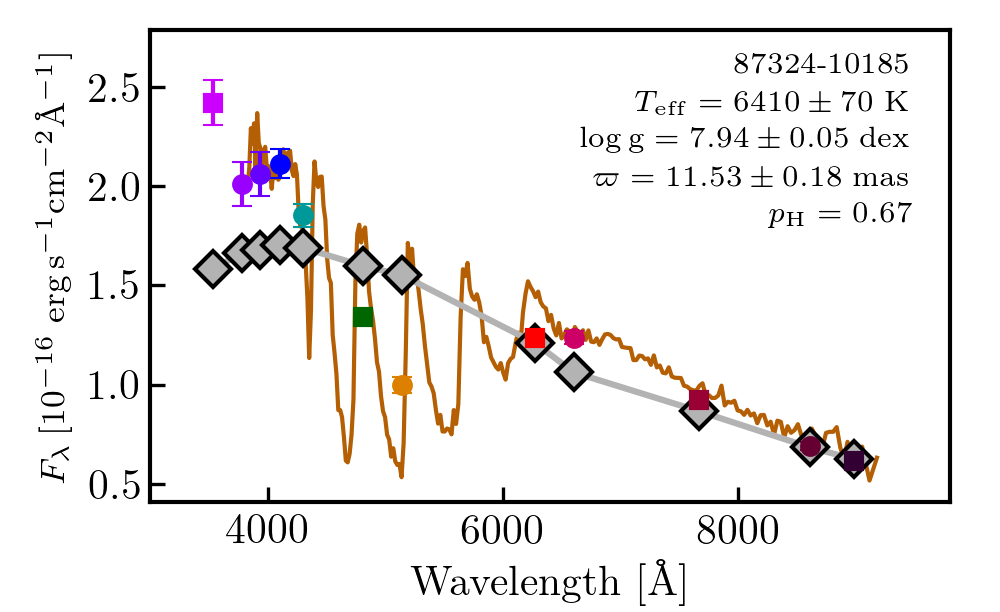}}
\caption{Same as Fig.~\ref{fig:examples}, but for two white dwarfs with $\tout = 3$, $M > 0.45$~$M_{\odot}$, and spectroscopically classified as DQ. The SDSS spectrum of the source is shown with the brown solid line.
}
\label{fig:dq}
\end{figure}

\begin{figure}[t]
\centering
\resizebox{\hsize}{!}{\includegraphics{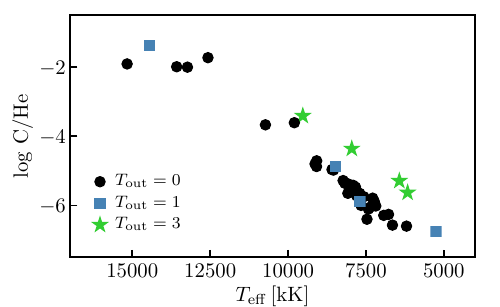}}
\caption{Carbon-to-helium abundance ratio (C/He) as a function of $\teff$ for DQs with spectroscopic measurements and $\tout = 0$ (black circles), $\tout = 1$ (blue squares), and $\tout = 3$ (green stars).
}
\label{fig:che}
\end{figure}

We identified a total of $46$ sources with $\tout = 3$ and $M > 0.45$~$M_{\odot}$ ($16$\% of the population). A detailed analysis using SIMBAD, the MWDD, and the available J-PLUS photometry reveals the following demographics: $18$ objects with contaminated photometry; 11 WDMS candidates (of which only two are spectroscopically confirmed); seven DQs (five with available spectra); five DAs (including a double degenerate, a confirmed ZZ Ceti, and two without spectra); one cataclysmic variable; one DO; one DBA; one DAB; and one DABH. Consequently, $63$\% of the $\tout = 3$ high-mass sources are classified as two-component systems, $33$\% as other noteworthy types, and only $4$\% appear to be standard non-variable DAs.

The identified DQs are cool and present prominent Swan molecular bands (Fig.~\ref{fig:dq}). Their residuals resemble the two-component behavior, with a clear deficit in $g$ and $J0515$. The C/He abundances and $\teff$ values available for DQs with spectra were retrieved from the MWDD, with $44$ sources in common with the J-PLUS DR3 catalog. Their C/He abundance as a function of $\teff$ is presented in Fig~\ref{fig:che}. Most DQs ($40$) have $\tout = 0$ or $\tout = 1$, consistent with random noise. The only four DQs with $\tout = 3$ fall within the sequence of DQs with strong carbon absorption, which typically have a $1$ dex larger abundance than normal DQs \citep{dufour05,koester06,coutu19,blouin19dq}. This implies that J-PLUS is sensitive to carbon absorption only when the spectrum is strongly affected. Therefore, a method to select DQ candidates with strong carbon features is to focus on sources having $\tout = 3$ and $M > 0.45$~$M_{\odot}$, with a $15$\% success rate after inspecting only $0.3$\% of the total initial sample.

In summary, $\tout = 3$ sources are primarily identified as two-component systems, either physically associated or projected. Additionally, DQs with strong carbon absorption are included in this category. The capabilities of the $\tout = 3$ population to select WDMS is further tested in Sect.~\ref{sec:wdms}.

\section{Comparison between J-PLUS and other classifications}\label{sec:class}
The comparison with SDSS DR16 spectroscopy, visually classified in GF21, is presented in \citet{clsj22pda}. In that work, the reliability of the probability $p_{\rm H}$ as an indicator of an H-dominated atmosphere is established by comparing the fraction of objects spectroscopically classified as DA as a function of $p_{\rm H}$. After applying the prior on spectral type (Sec.~\ref{sec:prior}), a one-to-one relation is obtained, demonstrating that the probabilities are reliable. A similar result was obtained in \citet{clsj24pca} for the probability $p_{\rm Ca}$ and the presence of metals in the atmosphere of white dwarfs. The reader is referred to these works for further details.

We repeated this analysis with the new J-PLUS DR3 catalog and obtained similar results. Therefore, in the following sections, the focus is on comparing the classifications obtained with J-PLUS DR3 to recent studies that have classified white dwarfs by applying machine learning techniques to SDSS DR17 spectra (\citealt{vincent23}; Sect.~\ref{sec:spec_sdss}), to the {\it Gaia} DR3 BP/RP spectral coefficients (\citealt{vincent24,garciazamora23,kao24}; Sect.~\ref{sec:XP}), and to visual classifications in DESI EDR (\citealt{desi_edr_wd}; Sect.~\ref{sec:spec_desi}). Finally, an evaluation of the performance in identifying WDMS systems is presented in Sect.~\ref{sec:wdms}.

\subsection{Comparison with spectroscopic classifications}\label{sec:spec_type}
In this section, the comparison with spectroscopic classifications was restricted to sources with $M \geq 0.45$~$M_{\odot}$ and $5\,000 < \teff < 30\,000$~K in J-PLUS DR3 to ensure adequate coverage of the theoretical models.

\begin{figure}[t]
\centering
\resizebox{\hsize}{!}{\includegraphics{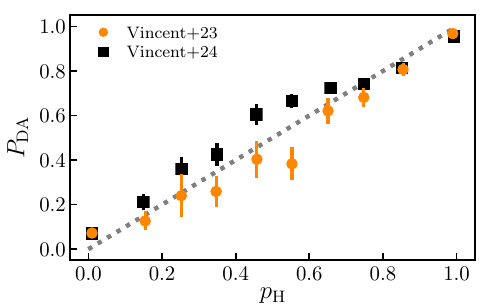}}
\resizebox{\hsize}{!}{\includegraphics{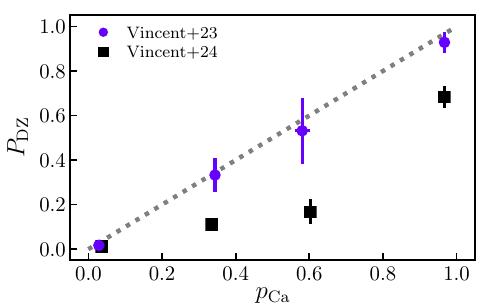}}
\caption{Mean probability from machine learning classification as a function of the mean probability from J-PLUS. Circles and squares show the results from SDSS DR17 spectra (\citealt{vincent23}, Sect.~\ref{sec:spec_sdss}) and {\it Gaia} BP/RP spectra (\citealt{vincent24}, Sect.~\ref{sec:spec_vin24}), respectively. The dotted line marks the one-to-one relation. {\it Top panel}: $P_{\rm DA}$ as a function of $p_{\rm H}$. {\it Bottom panel}: $P_{\rm DZ}$ as a function of $p_{\rm Ca}$.
}
\label{fig:pdx}
\end{figure}

\begin{figure}[t]
\centering
\resizebox{\hsize}{!}{\includegraphics{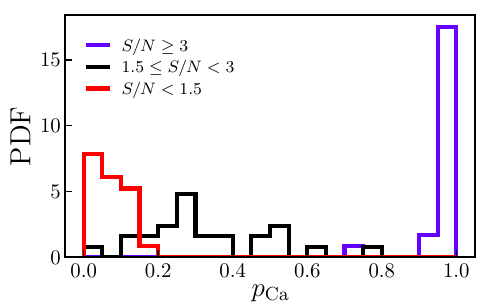}}
\caption{Probability distribution function of DZ white dwarfs from SDSS DR17 in $\pca$ for three different $S/N$ selections in ${\rm EW}_{J0395}$: $S/N \geq 3$ (purple), $1.5 \leq S/N < 3$ (black), and $S/N < 1.5$ (red).
}
\label{fig:hist_pca}
\end{figure}

\subsubsection{SDSS DR17 spectroscopy}\label{sec:spec_sdss}
In \citet{vincent23}, a total of $36\,523$ white dwarfs with spectra from SDSS DR17 ($R \sim 2\,000$) were analyzed using supervised machine learning tools. The classifier was trained on spectra with $S/N > 9$ and visual classifications from GF21. The model output corresponds to the probability of a given spectral type, with the primary type (DA, DB, DC, DZ, DQ, DO, among others) recovered with an accuracy of $\gtrsim 90$\%. 

The comparison was carried out on sources with $S/N > 9$ in SDSS DR17 spectra. A total of $2\,331$ objects were found in common. We assessed the reliability of $p_{\rm H}$ by comparing the mean $P_{\rm DA}$ from \citet{vincent23} with the mean $p_{\rm H}$ in J-PLUS, at different $p_{\rm H}$ values (Fig.~\ref{fig:pdx}). The results are consistent with the one-to-one relation, as expected, and deviate by less than $0.01$ from those obtained with the visual spectral classification from GF21. This outcome was anticipated given the performance of the \citet{vincent23} classifier.

Of the $72$ white dwarfs classified as DZ by \citet{vincent23} that are also present in our sample, only $29$ have $\pca > 0.5$. This can be explained as a reflection of the $S/N$ in ${\rm EW}_{J0395}$, which serves as the source of information on calcium absorption in J-PLUS. The distribution of $\pca$ for objects classified as DZ at different significance levels in ${\rm EW}_{J0395}$ is shown in Fig.~\ref{fig:hist_pca}. All sources with $S/N > 3$ have $\pca > 0.8$, while those with $S/N < 1.5$ have $\pca < 0.2$. As an additional test, the $P_{\rm DZ}$ probabilities were compared with $\pca$ (Fig.~\ref{fig:pdx}). Once again, the measurements are consistent with the one-to-one relation, demonstrating that the $\pca$ values accurately represent the capability of J-PLUS to detect the presence of calcium in white dwarf atmospheres.

\begin{figure}[t]
\centering
\resizebox{\hsize}{!}{\includegraphics{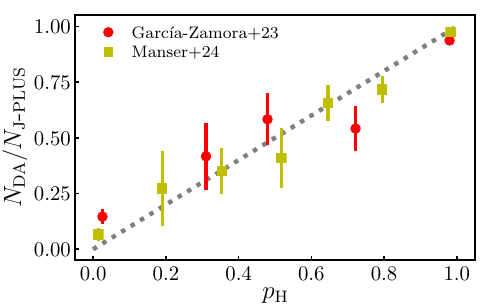}}
\caption{Fraction of white dwarfs classified as DA as a function of $p_{\rm H}$. Squares and circles show the results from DESI EDR (\citealt{desi_edr_wd}, Sect.~\ref{sec:spec_desi}) and {\it Gaia} BP/RP spectra (\citealt{garciazamora23}, Sect.~\ref{sec:spec_gz}), respectively. The dotted line marks the one-to-one relation.
}
\label{fig:nda}
\end{figure}

\subsubsection{DESI EDR spectroscopy}\label{sec:spec_desi}
The DESI EDR ($R \sim 3\,500$) provides visual spectral types for $2\,706$ individual white dwarfs and $66$ binary systems \citep{desi_edr_wd}. A total of $564$ objects were found in common with J-PLUS DR3. The fraction of DAs in DESI as a function of $p_{\rm H}$ is presented in Fig.~\ref{fig:nda}. The measurements are compatible with the one-to-one relation, providing another independent validation of the probabilities from J-PLUS. 

Only $15$ DESI DZs are included in the J-PLUS DR3 catalog, of which six have $\pca > 0.5$ or $\tout = 2$. As discussed in the previous section, this reflects the $S/N$ in ${\rm EW}_{J0395}$, with the six metal-polluted white dwarfs from J-PLUS having $S/N > 2.5$, while the remaining nine exhibit $S/N \lesssim 1$.

Comparison with spectral classification based on SDSS DR17 and DESI EDR confirmed that the probabilities $p_{\rm H}$ and $\pca$ are reliable, representing the correct fraction of DA and DZ spectral type objects, respectively. This allows these probabilities to be used for statistical studies, with appropriate weighting of each object \citep{clsj22pda}, to select DZ candidates with a high success rate \citep{clsj24pca}, and to test other classifications using J-PLUS as an intermediate step from spectroscopy. The latter case is explored in the following section, where the probabilities $p_{\rm H}$ and $\pca$ are compared with various classifications based on the BP/RP spectra from {\it Gaia} DR3.

\subsection{Gaia DR3 BP/RP spectra}\label{sec:XP}
In this section, classifications based on the {\it Gaia} DR3 BP/RP spectra ($R = 30-90$) are compared with the information in the J-PLUS DR3 catalog. The classifications from \citet{vincent24}, \citet{garciazamora23}, and \citet{kao24} are explored. In all cases, the $110$ coefficients that describe the BP/RP spectra were analyzed using machine learning techniques.

\subsubsection{Comparison with Vincent et al. 2024}\label{sec:spec_vin24}
Using a gradient-boosting classifier trained with the spectroscopic labels from SDSS DR17, \citet{vincent24} provide the main spectral-type (DA, DB, DC, DO, DZ, DQ) probability for $100\,188$ white dwarfs. Following the discussion in the previous section, the main probabilities for DA and DZ in \citet{vincent24} were compared with $p_{\rm H}$ and $\pca$ for the $5\,380$ objects in common (Fig.~\ref{fig:pdx}), after restricting the sample to sources with $M \geq 0.45$~$M_{\odot}$ and $5\,000 < \teff < 30\,000$~K in J-PLUS.

We find a close agreement between $P_{\rm DA}$ and $p_{\rm H}$, with a slight overconfidence for $P_{\rm DA}$ at $p_{\rm H} < 0.8$. The performance for DA classification reached $98$\% precision and recall, according to \citet{vincent24}. Therefore, the agreement with the J-PLUS probability is expected. The situation differs for metal-polluted systems, where we find a lower $P_{\rm DZ}$ with respect to $p_{\rm Ca}$. This suggests an under-confidence in the probability from \citet{vincent24}, reflecting the $60$\% precision and $85$\% recall estimated for their DZ classification.

\subsubsection{Comparison with Garc\'ia-Zamora et al. 2023}\label{sec:spec_gz}
A random forest algorithm is used by \citet{garciazamora23} to obtain the spectral types for $9\,448$ white dwarfs in the $100$ pc sample presented by \citet{jimenezesteban23}. In this case, sources with $M \geq 0.45$~$M_{\odot}$ and $6\,000 < \teff < 30\,000$~K in J-PLUS were used. A total of $465$ white dwarfs in common were found.

The fraction of white dwarfs classified as DA as a function of $p_{\rm H}$ is presented in Fig.~\ref{fig:nda}. The lower number of objects translates to large error bars, and the measurements are compatible with the desired one-to-one relation. The high precision and recall ($>90$\%) estimated by \citet{garciazamora23} for DAs are consistent with these results.

The number of metal-polluted white dwarfs is not large enough to perform the statistical analysis. On the one hand, there are $17$ DZs from \citet{garciazamora23} in the J-PLUS DR3 catalog, of which $15$ ($90$\%) have $\pca \geq 0.5$ or $\tout = 2$. The two discrepant sources have $S/N < 2.5$ in ${\rm EW}_{J0395}$ and are classified spectroscopically as DZs, highlighting the high precision ($>90$\%) of the \citet{garciazamora23} classification. On the other hand, there are $14$ sources with $\pca > 0.9$ in common, of which $12$ are classified as DZ, one as DA, and one as DC. This translates to an $85$\% agreement. There is no spectroscopic information for the two discrepant sources, but $J0395$ absorption is clearly detected in both systems and is even apparent in the {\it Gaia} BP/RP spectra. 

\begin{figure}[t]
\centering
\resizebox{\hsize}{!}{\includegraphics{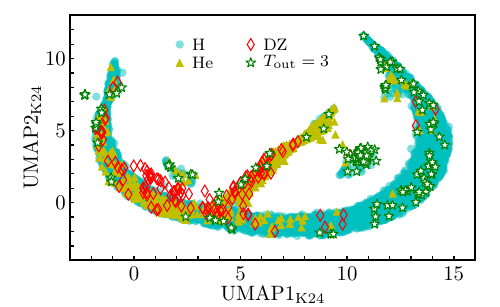}}
\caption{Manifold generated with \texttt{UMAP} from \citet{kao24}, based on {\it Gaia} BP/RP spectral coefficients. Symbols represent different J-PLUS classes: H-dominated atmosphere (cyan circles), He-dominated atmospheres (yellow triangles), metal-polluted atmospheres ($\pca > 0.5$ or $\tout = 2$; red diamonds), and $\tout = 3$ sources (green stars).}
\label{fig:kao}
\end{figure}

\subsubsection{Comparison with Kao et al. 2024}\label{sec:spec_kao}
As a final comparison, we studied the correlation between the results from J-PLUS and the \texttt{UMAP} manifold analysis presented in \citet{kao24}, hereafter denoted UMAP$_{\rm K24}$. A total of $6\,323$ sources are found in common (Fig.~\ref{fig:kao}). We obtained results similar to those reported in \citet{kao24} using spectroscopic classes. The UMAP$_{\rm K24}$ manifold shown in Fig.~\ref{fig:kao} exhibits a characteristic horseshoe shape, with H-dominated white dwarfs forming the majority of its population. This structure reflects an effective-temperature sequence in which the hotter DAs are located on the right side of the horseshoe, with progressively cooler ones extending toward the left. The central arm is occupied by He-dominated atmospheres. Metal-polluted systems with $\pca > 0.5$ or $\tout = 2$ dominate the region at UMAP$_{\rm K24}$ coordinates (1,2), and are also spread along the He-dominated arm and at cool temperatures, consistent with the results of \citet{kao24}. Finally, the $\tout = 3$ sources are highlighted. The population at UMAP$_{\rm K24}$ coordinates (2,2), suggested as binaries by \citet{kao24}, is not well populated by $\tout = 3$ sources. An inspection shows that half of all the white dwarfs in this area ($85$\% for those with $\tout = 3$) are associated in J-PLUS images with a nearby red source, which is likely contaminating the BP/RP spectra at the red end. A concentration is observed at UMAP$_{\rm K24}$ coordinates (11,3). In this case, only $20$\% of the $\tout = 3$ sources have contaminated photometry, suggesting that this area is dominated by unresolved binary systems (see \citealt{perezcouto25} for a similar conclusion). The location of DQs with strong carbon absorption was also explored. An excess is found at UMAP$_{\rm K24}$ coordinates (-2.3,7.5), where a strip is present in the manifold of \citet{kao24}. This was not discussed by the authors, and the sources in this strip are classified as DQs by \citet{vincent24}.

\subsection{Selection of WDMS systems}\label{sec:wdms}
In this section, we compare several spectroscopic catalogs of WDMS systems with the classification from J-PLUS, particularly focusing on sources with $\tout = 3$, as they are dominated by two-component systems. All sources in the J-PLUS DR3 white dwarf catalog were used in the comparison. 

The first comparison is with the WDMS catalog from SDSS spectra collected over multiple releases \citep{wdms_sdss_i,wdms_sdss_ii,wdms_sdss_iii}. The latest version of the catalog contains $3287$ systems from SDSS DR12, with $45$ in common with J-PLUS DR3. Among these sources, $26$ ($58$\%) are assigned $\tout = 3$, while $19$ (42\%) are assigned $\tout = 1$ or $0$. When a mass-based selection is applied, objects with $\tout = 3$ account for $77$\% (24/31) and $14$\% (2/14) of the sources with masses below and above $0.45\,M_\odot$, respectively.

A machine learning model was trained by \citet{vincent23} to identify binary sources in SDSS DR17 spectra. In the total J-PLUS DR3 catalog, $82$ objects were classified as WDMS by \citet{vincent23}, with $46, 6, 0,$ and $30$ assigned to $\tout = 0, 1, 2,$ and $3$, respectively. The outlier type does not appear to be effective in selecting binary systems, as only 36\% are placed in the two-component class, while 65\% are assigned to other categories. However, the $\tout = 3$ sources account for $90$\% (28/31) of spectroscopic binaries with masses below $0.45\,M_{\odot}$ and $4$\% (2/51) of those above this limit. 

Seven WDMS from DESI \citep{desi_edr_wd} are classified in the J-PLUS catalog, with six having $M \leq 0.45$~$M_{\odot}$. Of the low-mass objects, five ($85$\%) are assigned $\tout = 3$. The single object with $M > 0.45$~$M_{\odot}$ has $\tout = 0$. 

Finally, a total of $14$ WDMS are in common between J-PLUS DR3 and the Large Sky Area Multi-Object Fiber Spectroscopic Telescope (LAMOST, \citealt{lamost}) DR5 binary catalog presented in \citet{wdms_lamost_ii}. In this case, all seven low-mass objects with $M \leq 0.45\,M_{\odot}$ are classified as $\tout = 3$ sources, while 20\% (2/7) of the $M > 0.45\,M_{\odot}$ sources have  $\tout = 3$.

In summary, the comparison with spectroscopic WDMS catalogs shows that $80-100$\% of the WDMS systems with low-mass solutions in J-PLUS have $\tout = 3$, compared to only $5-20$\% of the WDMS systems with $M > 0.45$~$M_{\odot}$. This result suggests that objects with clearly distorted colors compared to the case of a single white dwarf are classified as $\tout = 3$, which in most cases also implies a brighter magnitude (lower mass) than expected for a single object. Cases with a subtle companion exhibit both J-PLUS colors and magnitudes compatible with the single white dwarf scenario.

\section{Summary and conclusions}\label{sec:conclusion}
The J-PLUS DR3 white dwarf catalog comprises a total of $\nwd$ sources with $r \leq 20$ mag and parallaxes in the range $1 \leq \varpi < 100$ mas over $3\,284$ deg$^2$. We anallysed the $12$-band J-PLUS photometry with Bayesian SED-fitting techniques to derive the effective temperature and surface gravity of the sources. Priors on parallax and spectral type were applied to improve the reliability of the estimates. Furthermore, we determined the probability that each object has an H-dominated atmosphere ($p_{\rm H}$) and exhibiting calcium absorption ($p_{\rm Ca}$).

We applied the \texttt{UMAP} dimensionality reduction technique to the $12$-dimensional, uncertainty-normalized residuals from the SED-fitting analysis to investigate outliers, defined as sources that exhibit significant deviations from theoretical atmospheric models ($\chi^2 \geq 23.2$; 771 sources). The resulting 2D distribution reveals three distinct types of outliers: $T_{\rm out} = 1$ (391 sources), compatible with random measurement fluctuations and occupying the same \texttt{UMAP} locus as sources with $\chi^2 < 23.2$; $T_{\rm out} = 2$ (98 sources), comprising metal-polluted white dwarfs; and $T_{\rm out} = 3$ (282 sources), dominated by two-component systems consisting of a white dwarf with a red companion (physical or in projection), and also including DQs with strong carbon absorption features.

The J-PLUS probabilities were compared with atmospheric types from SDSS DR17, DESI EDR, and \textit{Gaia} DR3 BP/RP spectra. We conclude that both $p_{\rm H}$ and $p_{\rm Ca}$ are reliable, accurately representing the probability of selecting a DA and a DZ, respectively.

The J-PLUS DR3 catalog presented in this work serves as a starting point for future statistical analyses of the white dwarf population. The techniques developed for the analysis of outliers will be essential for the ongoing J-PAS project. With its $56$ narrow optical bands, J-PAS will enhance spectral classification \citep{clsj22mjpaswd} and improve the characterization of outliers with respect to J-PLUS.

\section*{Data availability}
The J-PLUS DR3 white dwarf catalog described in Appendix~\ref{app:catalog} is accessible at the CDS via \url{https://cdsarc.cds.unistra.fr/viz-bin/cat/J/A+A/701/A273}. It is also accessible in the J-PLUS database\footnote{\url{https://archive.cefca.es/catalogues/jplus-dr3}} at the table \texttt{jplus.WhiteDwarf}, and through the Virtual Observatory Table Access Protocol (TAP) service\footnote{\url{https://archive.cefca.es/catalogues/vo/tap/jplus-dr3}}.

\begin{acknowledgements}
We dedicate this paper to the memory of our six IAC colleagues and friends who met with a fatal accident in Piedra de los Cochinos, Tenerife, in February 2007, with  special thanks to Maurizio Panniello, whose teachings of \texttt{python} were so important for this paper.

We thank the anonymous referee for useful comments and suggestions.

Based on observations made with the JAST80 telescope at the Observatorio Astrof\'{\i}sico de Javalambre (OAJ), in Teruel, owned, managed, and operated by the Centro de Estudios de F\'{\i}sica del  Cosmos de Arag\'on. We acknowledge the OAJ Data Processing and Archiving Unit (UPAD, \citealt{upad}) for reducing and calibrating the OAJ data used in this work.

Funding for the J-PLUS Project has been provided by the Governments of Spain and Arag\'on through the Fondo de Inversiones de Teruel; the Aragonese Government through the Research Groups E96, E103, E16\_17R, E16\_20R, and E16\_23R; the Spanish Ministry of Science and Innovation (MCIN/AEI/10.13039/501100011033 y FEDER, Una manera de hacer Europa) with grants PID2021-124918NB-C41, PID2021-124918NB-C42, PID2021-124918NA-C43, and PID2021-124918NB-C44; the Spanish Ministry of Science, Innovation and Universities (MCIU/AEI/FEDER, UE) with grants PGC2018-097585-B-C21 and PGC2018-097585-B-C22; the Spanish Ministry of Economy and Competitiveness (MINECO) under AYA2015-66211-C2-1-P, AYA2015-66211-C2-2, AYA2012-30789, and ICTS-2009-14; and European FEDER funding (FCDD10-4E-867, FCDD13-4E-2685). The Brazilian agencies FINEP, FAPESP, and the National Observatory of Brazil have also contributed to this project.

This research received funding from the European Research Council under the European Union’s Horizon 2020 research and innovation programme number 101002408 (MOS100PC).

A.~d.~P. acknowledges financial support from the Severo Ochoa grant CEX2021-001131-S funded by MCIN/AEI/10.13039/501100011033.

A.~E., J.~A.~F.~O., and J.~V.~F. acknowledge the financial support from the Spanish Ministry of Science and Innovation and the European Union - NextGenerationEU through the Recovery and Resilience Facility project ICTS-MRR-2021-03-CEFCA.

This work was partially supported by MICIU/AEI/10.13039/501100011033 grants PID2023-148661NB-I00, PID2023-146210NB-I00 and the AGAUR/Generalitat de Catalunya grant SGR-386/2021.

P.~C. acknowledges financial support from the Spanish Ministry of Science and Innovation/State Agency of Research MCIN/AEI/10.13039/501100011033 through grants PID2019-107061GB-C61 and the Spanish Virtual Observatory project PID2020-112949GB-I00.

J.~V. acknowledges the technical members of the UPAD for their invaluable work: Juan Castillo, Javier Hern\'andez, \'Angel L\'opez, Alberto Moreno, and David Muniesa.

This work has made use of data from the European Space Agency (ESA) mission
{\it Gaia} (\url{https://www.cosmos.esa.int/gaia}), processed by the {\it Gaia} Data Processing and Analysis Consortium (DPAC, \url{https://www.cosmos.esa.int/web/gaia/dpac/consortium}). Funding for the DPAC has been provided by national institutions, in particular the institutions participating in the {\it Gaia} Multilateral Agreement.

Funding for SDSS-III has been provided by the Alfred P. Sloan Foundation, the Participating Institutions, the National Science Foundation, and the U.S. Department of Energy Office of Science. The SDSS-III web site is \url{http://www.sdss3.org/}.

SDSS-III is managed by the Astrophysical Research Consortium for the Participating Institutions of the SDSS-III Collaboration including the University of Arizona, the Brazilian Participation Group, Brookhaven National Laboratory, Carnegie Mellon University, University of Florida, the French Participation Group, the German Participation Group, Harvard University, the Instituto de Astrofisica de Canarias, the Michigan State/Notre Dame/JINA Participation Group, Johns Hopkins University, Lawrence Berkeley National Laboratory, Max Planck Institute for Astrophysics, Max Planck Institute for Extraterrestrial Physics, New Mexico State University, New York University, Ohio State University, Pennsylvania State University, University of Portsmouth, Princeton University, the Spanish Participation Group, University of Tokyo, University of Utah, Vanderbilt University, University of Virginia, University of Washington, and Yale University.

Funding for the Sloan Digital Sky Survey IV has been provided by the  Alfred P. Sloan Foundation, the U.S. Department of Energy Office of Science, and the Participating  Institutions. SDSS-IV acknowledges support and resources from the Center for High Performance Computing  at the  University of Utah. The SDSS  website is \url{www.sdss.org}. SDSS-IV is managed by the Astrophysical Research Consortium for the Participating Institutions of the SDSS Collaboration including the Brazilian Participation Group, the Carnegie Institution for Science, Carnegie Mellon University, Center for Astrophysics | Harvard \& Smithsonian, the Chilean Participation Group, the French Participation Group, Instituto de Astrof\'isica de Canarias, The Johns Hopkins University, Kavli Institute for the Physics and Mathematics of the Universe (IPMU) / University of Tokyo, the Korean Participation Group, Lawrence Berkeley National Laboratory, Leibniz Institut f\"ur Astrophysik Potsdam (AIP),  Max-Planck-Institut f\"ur Astronomie (MPIA Heidelberg), Max-Planck-Institut f\"ur Astrophysik (MPA Garching), Max-Planck-Institut f\"ur Extraterrestrische Physik (MPE), National Astronomical Observatories of China, New Mexico State University, New York University, University of Notre Dame, Observat\'ario Nacional / MCTI, The Ohio State University, Pennsylvania State University, Shanghai Astronomical Observatory, United Kingdom Participation Group, Universidad Nacional Aut\'onoma de M\'exico, University of Arizona, University of Colorado Boulder, University of Oxford, University of Portsmouth, University of Utah, University of Virginia, University of Washington, University of Wisconsin, Vanderbilt University, and Yale University.

DESI construction and operations is managed by the Lawrence Berkeley National Laboratory. This research is supported by the U.S. Department of Energy, Office of Science, Office of High-Energy Physics, under Contract No. DE–AC02–05CH11231, and by the National Energy Research Scientific Computing Center, a DOE Office of Science User Facility under the same contract. Additional support for DESI is provided by the U.S. National Science Foundation, Division of Astronomical Sciences under Contract No. AST-0950945 to the NSF’s National Optical-Infrared Astronomy Research Laboratory; the Science and Technology Facilities Council of the United Kingdom; the Gordon and Betty Moore Foundation; the Heising-Simons Foundation; the French Alternative Energies and Atomic Energy Commission (CEA); the National Council of Science and Technology of Mexico (CONACYT); the Ministry of Science and Innovation of Spain, and by the DESI Member Institutions. The DESI collaboration is honored to be permitted to conduct astronomical research on Iolkam Du’ag (Kitt Peak), a mountain with particular significance to the Tohono O’odham Nation.

This research has made use of the SIMBAD database, operated at CDS, Strasbourg, France.

This research made use of \texttt{Astropy}, a community-developed core \texttt{Python} package for Astronomy \citep{astropy}, and \texttt{Matplotlib}, a 2D graphics package used for \texttt{Python} for publication-quality image generation across user interfaces and operating systems \citep{pylab}. IA generative models have been used to revise the text.
\end{acknowledgements}

\bibliographystyle{aa}
\bibliography{biblio}

\newpage

\begin{appendix}
\onecolumn
\section{Description of the J-PLUS DR3 white dwarf catalog}\label{app:catalog}
The description of the columns in the catalog is provided in Table.~\ref{tab:catalog}. 
\begin{table*}[h] 
\caption{J-PLUS DR3 white dwarf catalog.}
\label{tab:catalog}
\centering 
        \begin{tabular}{c c l}
        \hline\hline\rule{0pt}{3ex} 
        Heading     & Units  &  Description   \\
        \hline\rule{0pt}{3ex}
      \!TileID     &  $\cdots$       & Identifier of the J-PLUS Tile image in the $r$ band where the object was detected. \\ 
        Number       &  $\cdots$       & Number identifier assigned by \texttt{SExtractor} for the object in the $r$ band image. \\
        RAJ2000        &  deg            & Right ascension (J2000). \\
        DEJ2000        &  deg            & Declination (J2000). \\
        rmag         &  mag            & De-reddened total $r-$band apparent magnitude, noted $\hat{r}$.     \\ 
        e\_rmag      &  mag            & Uncertainty in $\hat{r}$.\\ 
        psel        &  $\cdots$       & Selection probability for $\hat{r} \leq 20$ mag and $1 \leq \varpi \leq 100$ mas.\\ 
        pH          &  $\cdots$       & Probability of having a H-dominated atmosphere.\\ 
        TeffH      &  K              & Effective temperature for a H-dominated atmosphere. \\ 
        e\_TeffH   &  K              & Uncertainty in TeffH.\\
        loggH      &  dex            & Decimal logarithm of the surface gravity for a H-dominated atmosphere. \\ 
        e\_loggH   &  dex            & Uncertainty in loggH.\\
        plxH       &  mas            & Parallax for a H-dominated atmosphere. \\ 
        e\_plxH    &  mas            & Uncertainty in plxH.\\
        massH      &  $M_{\odot}$    & Mass for a H-dominated atmosphere. \\ 
        e\_massH   &  $M_{\odot}$    & Uncertainty in massH.\\ 
        VH      &  ${\rm pc}^3$      & Effective volume at $10 \leq d \leq 1000$ pc for a H-dominated atmosphere. \\ 
        TeffHe     &  K              & Effective temperature for a He-dominated atmosphere. \\ 
        e\_TeffHe  &  K              & Uncertainty in TeffHe.\\
        loggHe     &  dex            & Decimal logarithm of the surface gravity for a He-dominated atmosphere. \\ 
        e\_loggHe  &  dex            & Uncertainty in loggHe.\\
        plxHe      &  mas            & Parallax for a He-dominated atmosphere. \\ 
        e\_plxHe   &  mas            & Uncertainty in plxHe.\\
        massHe     &  $M_{\odot}$    & Mass for a He-dominated atmosphere. \\ 
        e\_massHe  &  $M_{\odot}$    & Uncertainty in massHe.\\ 
        VHe        &  ${\rm pc}^3$   & Effective volume at $10 \leq d \leq 1000$ pc for a He-dominated atmosphere. \\ 
        pca        &  $\cdots$       & Probability of having \ion{Ca}{ii} H+K absorption.\\
        EWJ0395    & \AA             & EW of the $J0395$ passband.\\
        e\_EWJ0395 & \AA   & Error in EWJ0395.\\
        TeffCa     &  K              & Effective temperature from the calcium absorption analysis.\\ 
        e\_TeffCa  &  K              & Uncertainty in TeffCa.\\
        loggCa     &  dex            & Decimal logarithm of the surface gravity from the calcium absorption analysis. \\ 
        e\_loggCa  &  dex            & Uncertainty in loggCa.\\
        plxCa      &  mas            & Parallax from the calcium absorption analysis. \\ 
        e\_plxCa   &  mas            & Uncertainty in plxCa.\\
        massCa     &  $M_{\odot}$    & Mass from the calcium absorption analysis. \\ 
        e\_massCa  &  $M_{\odot}$    & Uncertainty in massCa.\\ 
        chi        &  $\cdots$       & Uncertainty-normalized residuals from the best-fitting model without the $J0395$ filter\\
        chi2       &  $\cdots$       & $\chi^2$ from the best-fitting model\\
        umap1      &  $\cdots$       & First dimension of the \texttt{UMAP} manifold\\
        umap2      &  $\cdots$       & Second dimension of the \texttt{UMAP} manifold\\
        Typeout       &  $\cdots$       & Outlier type\\
        \hline 
\end{tabular}
\end{table*}

\section{Median residuals along the \texttt{UMAP} manifold}\label{app:chi_umap}
\begin{figure*}[t]
\centering
\resizebox{0.33\hsize}{!}{\includegraphics{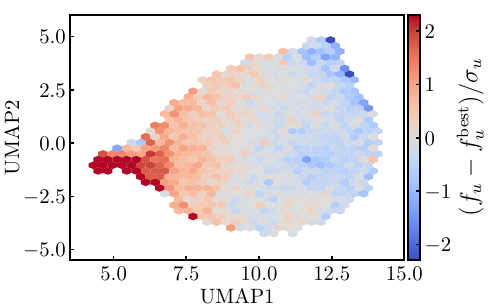}}
\resizebox{0.33\hsize}{!}{\includegraphics{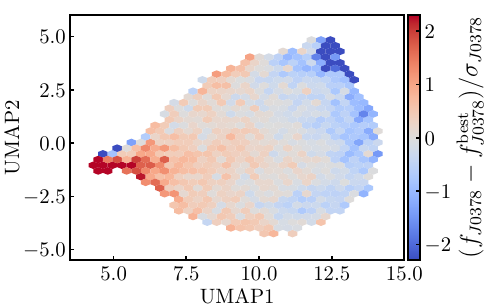}}
\resizebox{0.33\hsize}{!}{\includegraphics{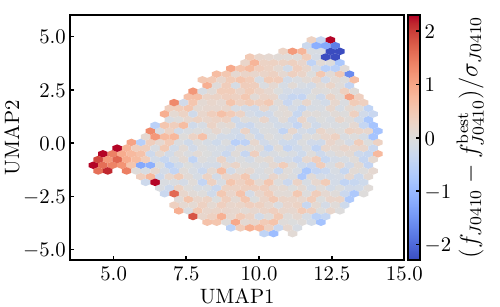}}\\
\resizebox{0.33\hsize}{!}{\includegraphics{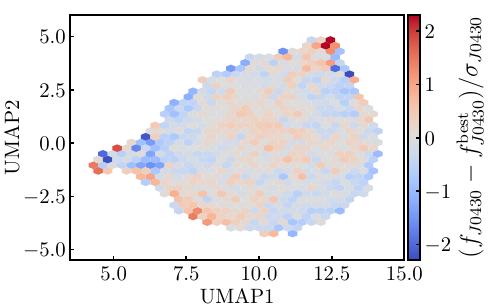}}
\resizebox{0.33\hsize}{!}{\includegraphics{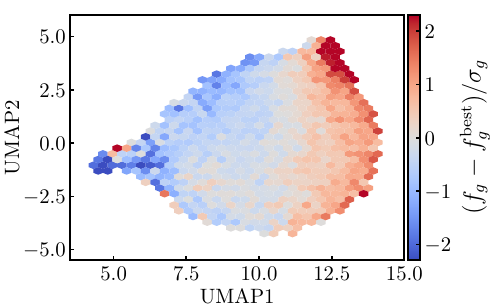}}
\resizebox{0.33\hsize}{!}{\includegraphics{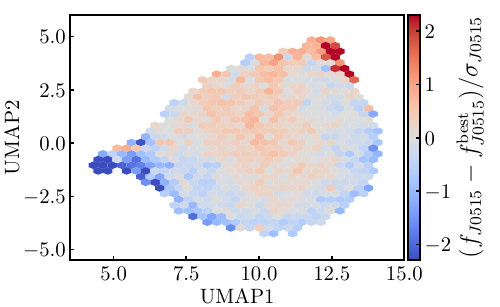}}\\
\resizebox{0.33\hsize}{!}{\includegraphics{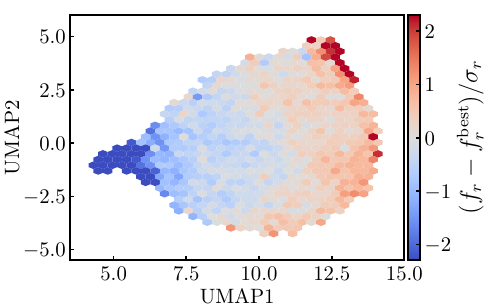}}
\resizebox{0.33\hsize}{!}{\includegraphics{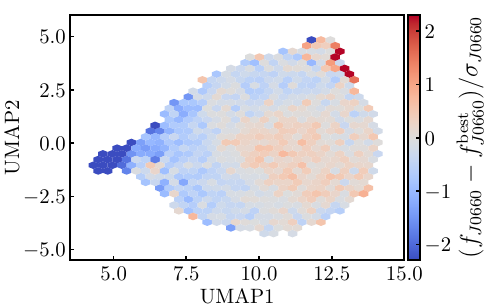}}
\resizebox{0.33\hsize}{!}{\includegraphics{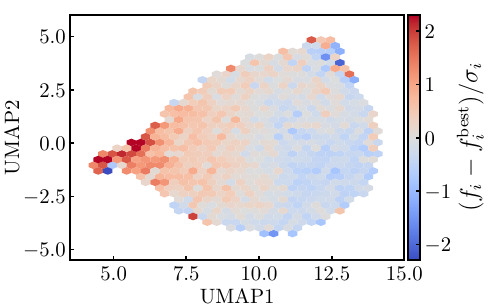}}\\
\resizebox{0.33\hsize}{!}{\includegraphics{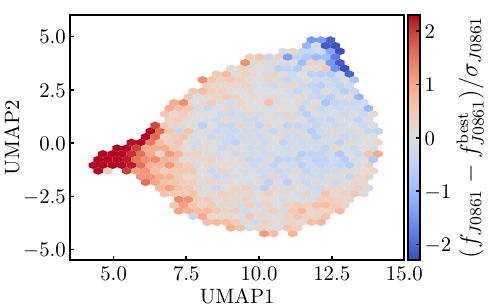}}
\resizebox{0.33\hsize}{!}{\includegraphics{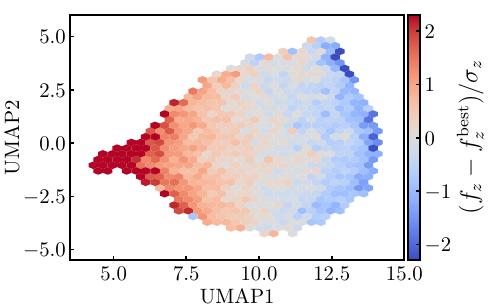}}
\caption{Median value of the normalized residuals along the \texttt{UMAP} two-dimensional manifold for the $u$, $J0378$, $J0410$, $J0430$, $g$, $J0515$, $r$, $J0660$, $i$, $J0861$, and $z$ passbands, from top to bottom and left to right.
}
\label{fig:chi_umap}
\end{figure*}

The median normalized residuals along the \texttt{UMAP} two-dimensional manifold for the J-PLUS passbands but $J0395$, already presented in Fig.~\ref{fig:chi_umap_j0395}, are shown in Fig.~\ref{fig:chi_umap}.

\end{appendix}

\end{document}